\documentclass[12pt, draftclsnofoot, onecolumn]{IEEEtran}
\newtheorem{definition}{Definition}
\newtheorem{lemma}{Lemma}
\newtheorem{theorem}{Theorem}
\newtheorem{remark}{Remark}
\newtheorem{corollary}{Corollary}
\usepackage{float}
\usepackage{graphicx}
\usepackage{enumerate}
\usepackage{amsmath}
\usepackage{amssymb}
\usepackage{amsfonts}
\usepackage{filecontents}
\usepackage{xcolor}
\usepackage{bm}
\usepackage{cite}
\usepackage{breqn}
\usepackage[colorlinks,linkcolor=blue,citecolor=blue]{hyperref}

\ifCLASSOPTIONcompsoc
\usepackage[caption=false,font=normalsize,labelfon
t=sf,textfont=sf]{subfig}
\else
\usepackage[caption=false,font=footnotesize]{subfi
	g}
\fi

\begin{document}
\title{Computation over Wide-Band MAC: Improved Achievable Rate through Sub-Function Allocation}

\author{Fangzhou Wu, Li Chen, Nan Zhao, \IEEEmembership{Senior Member, IEEE,} Yunfei Chen, \IEEEmembership{Senior Member, IEEE,} F. Richard Yu, \IEEEmembership{Fellow, IEEE,} and Guo Wei
	\thanks{F. Wu, L. Chen and G. Wei are with Department of Electronic Engineering and Information Science, University of Science and Technology of China, Hefei, Anhui 230027. (e-mail: fangzhouwu@outlook.com, \{chenli87, wei\}@ustc.edu.cn).}
	\thanks{N. Zhao is with the School of Info. and Commun. Eng., Dalian University of Technology, Dalian 116024, China, and also with National Mobile Communications Research Laboratory, Southeast University, Nanjing 210096, China. (e-mail:zhaonan@dlut.edu.cn).}
	\thanks{Y. Chen is with the School of Engineering, University of Warwick, Coventry CV4 7AL, U.K. (e-mail: Yunfei.Chen@warwick.ac.uk).}
	\thanks{F.R. Yu is with the Department of Systems and Computer Engineering, Carleton University, Ottawa, ON, K1S 5B6, Canada. (email: richard.yu@carleton.ca).}
}
\maketitle

\begin{abstract}
	Future networks are expected to connect an enormous number of nodes wirelessly using wide-band transmission. This brings great challenges. To avoid collecting a large amount of data from the massive number of nodes, computation over multi-access channel (CoMAC) is proposed to compute a desired function over the air utilizing the signal-superposition property of MAC. Due to frequency selective fading, wide-band CoMAC is more challenging and has never been studied before. In this work, we propose the use of orthogonal frequency division multiplexing (OFDM) in wide-band CoMAC to transmit functions in a similar way to bit sequences through division, allocation and reconstruction of function. An achievable rate without any adaptive resource allocation is derived. To prevent a vanishing computation rate from the increase of the number of nodes, a novel sub-function allocation of sub-carriers is derived. Furthermore, we formulate an optimization problem considering power allocation. A sponge-squeezing algorithm adapted from the classical water-filling algorithm is proposed to solve the optimal power allocation problem. The improved computation rate of the proposed framework and the corresponding allocation has been verified through both theoretical analysis and simulation.
\end{abstract}


\section{Introduction}
\IEEEPARstart{C}{urrent} 5G and internet of things envision the interconnections of up to 1 trillion products, machines and devices through wide-band transmission\cite{fettweis20145g,al2015internet,gupta2015survey}. With such an enormous number of nodes, it is impractical to transmit data using conventional orthogonal multi-access schemes for the future networks because this would result in excessive network latency and low spectrum utilization efficiency.

To solve this problem, computation over multi-access channel (CoMAC) was proposed to exploit the signal-superposition property of MAC to compute the desired function through concurrent node transmissions that combines computation with communication\cite{goldenbaum2015nomographic,goldenbaum2013robust,abari2016over,goldenbaum2014channel,kortke2014analog,zhu2018over,wu2019experimental,nazer2007computation,appuswamy2014computing,erez2005lattices,wu2018experimental,nazer2011compute,jeon2014computation,Jeon2016Opportunistic,wang2015interactive,goldenbaum2014computation}. A straightforward application of CoMAC to attain transmission and computation in parallel is wireless sensor networks whose goal of communication is typically for a fusion center to obtain a function value of the sensor readings (e.g., arithmetic mean, polynomial or number of active nodes), rather than store all readings from all sensors. Apart from this, CoMAC can also be applied in networks that focus on computing a class of so called nomographic functions of distributed data via concurrent node transmissions\cite{goldenbaum2015nomographic}. 

A simple computation framework, called analog CoMAC, has been studied in \cite{goldenbaum2013robust,abari2016over,goldenbaum2014channel,kortke2014analog,zhu2018over,wu2019experimental}. In analog CoMAC, all nodes participate in the transmission. The authors used pre-processing at each node and post-processing at the fusion center to deal with fading and compute or estimate functions\cite{goldenbaum2013robust}. The designs of pre-processing and post-processing were used to compute linear and non-linear functions discussed in\cite{abari2016over}, and the effect of channel estimation error was characterized in \cite{goldenbaum2014channel}. In order to verify whether the analog CoMAC is feasible in practice, an implementation using software defined radio was built in \cite{kortke2014analog}. A multi-function computation method utilizing a multi-antenna fusion center to collect data transmitted by a cluster of multi-antenna multi-modal sensors has been presented in \cite{zhu2018over}. Besides, using the multiplexing gain to compute multi-function has been discussed in \cite{wu2019experimental}. In summary, the simple analog CoMAC has led to an active area focusing on the design and implementation techniques for receiving a desired function of concurrent signals.

The main limitation of analog CoMAC is the lack of robust design against noise. Hence, digital CoMAC was proposed to use joint source-channel coding in \cite{nazer2007computation,appuswamy2014computing,erez2005lattices,nazer2011compute,jeon2014computation,Jeon2016Opportunistic,wang2015interactive,goldenbaum2014computation,wu2018experimental} to reduce the noise effect. The potential of linear source coding was discussed in \cite{nazer2007computation}, and its application was presented in \cite{appuswamy2014computing} for the function computation over MAC. Compared with linear source coding, nested lattice coding can approach the performance of standard random coding \cite{erez2005lattices}. This lattice-based CoMAC was extended to a general framework in \cite{nazer2011compute} for relay networks with linear channels and additive white Gaussian noise. In order to combat non-uniform fading, a uniform-forcing transceiver design was given in \cite{wu2018experimental}. Available computation rates were given in \cite{nazer2011compute,jeon2014computation,Jeon2016Opportunistic} for digital CoMAC through theoretical analysis. In \cite{Jeon2016Opportunistic}, Jeon and Bang found that the computation rates achieved by the above computation techniques decrease as the number of nodes increases, and eventually go to zero due to fading MAC. To handle this problem, they proposed an opportunistic CoMAC where a subset of nodes opportunistically participate in the transmission at each time, which provided a non-vanishing computation rate even when the number of nodes in the network goes to infinity.

To the best of our knowledge, the above mentioned works only consider a narrow-band CoMAC. A wide-band CoMAC has not been discussed before. Wide-band signal transmission has to face frequency selective fading because the coherence bandwidth of the channel is smaller than the signal bandwidth. In the time domain, the multi-path interference is not easy to remove\cite{tse2005fundamentals}. Orthogonal frequency division multiplexing (OFDM) is an effective solution to inter-symbol interference caused by a dispersive channel\cite{hasan2009energy,mohanram2007joint,chen2008fast}, and has been applied in many communication fields.

Motivated by the above observations, we employ OFDM in the implementation of CoMAC. However, the conventional OFDM systems which load bit sequences of different lengths into different sub-carriers cannot be used directly because CoMAC transmits a desired function over the air, not a bit sequence. Thus, the desired function as a whole cannot be allocated into several sub-carriers. To deal with the issue, we propose a method of sub-function allocation for the division, allocation and reconstruction of the desired function. This method divides the desired function to some sub-functions, allocates these sub-functions into different sub-carriers, and reconstructs the desired function at the fusion center. The theoretical expression of achievable computation rate is derived based on the classical results of nested lattice coding\cite{jeon2014computation,Jeon2016Opportunistic,wang2015interactive,goldenbaum2014computation}. It is shown to provide a non-vanishing computation rate as the number of nodes increases. An optimization considering power allocation is further discussed, and a sponge-squeezing algorithm adapted from the classical water-filling algorithm is proposed to solve the optimal power allocation problem. Owing to sub-function allocation and power allocation, the computation rate is improved significantly. Our contributions can be summarized as follows: 
\begin{itemize}
	\item \emph{Novel CoMAC-OFDM}. In order to tackle frequency selective fading, we exploit the mechanism of OFDM to implement wide-band CoMAC. Unlike the conventional CoMAC schemes, CoMAC-OFDM divides a desired function to sub-functions, allocates sub-functions to each sub-carrier and reconstructs the desired function at the fusion center.
	\item \emph{Improved computation rate}. A novel sub-function allocation is designed to assign sub-functions into sub-carriers, which provides a non-vanishing computation rate as the number of nodes increases. After that, the theoretical expression of improved computation rate is derived.
	\item \emph{Optimal power allocation}. An optimal power allocation is considered to improve computation rate. Since the solution of our optimization is not suitable to utilize the classical water-filling algorithm, we propose a sponge-squeezing algorithm adapted from the classical water-filling algorithm to carry out the power allocation, which provides a simple and concise interpretation of the necessary optimality conditions.
\end{itemize}

The paper is organized as follows. Section \ref{Preliminaries Network Model} introduces the definitions of CoMAC and the system model for CoMAC-OFDM. In Section \ref{Problem Statement and Main Results}, we summarize the main results of this paper and compare them with the existing results. Section \ref{Sub-Carrier Loading based on OFDM-INC} presents the proposed CoMAC-OFDM with sub-function allocation in detail and analyzes the computation rate. Section \ref{Power Allocation Methods for Sub-Carrier Loading} focuses on the improvement of the computation rate using power allocation. Simulation results are presented in Section \ref{Numerical Results} and conclusions are given in Section \ref{Conclusion}.
\section{Preliminaries \& Network Model}
\label{Preliminaries Network Model}
In this section, we give some definitions before introducing CoMAC. Throughout this paper, we denote $ [1:n]=\left\lbrace1,2,\cdots,n\right\rbrace$, and $ \mathsf{C}^+(x)=\max\left\lbrace \frac{1}{2}\log(x),0\right\rbrace$. For a set $ \mathcal{A} $, $ \left|\mathcal{A}\right|  $  denotes the cardinality of $ \mathcal{A} $. $ \lceil  x \rceil $ is the ceiling function as $ \lceil  x \rceil=\min\left\lbrace n\in\mathbb{Z}|x\le n \right\rbrace  $. Let the entropy of a random variable $ A $ be $ H(A) $ and $ \mathrm{diag}\left\lbrace a_1, a_2, \cdots, a_n\right\rbrace  $ denote the diagonal matrix of which the diagonal elements are from $ a_1 $ to $ a_n $. A set $ \left\lbrace x_1, x_2, \cdots, x_N \right\rbrace  $ is written as $ \left\lbrace x_i \right\rbrace_{i\in[1:N]}  $ for short. $ (\cdot)^{\mathrm{T}} $ represents the transpose of a vector or matrix.

\begin{figure}
	\centering
	\hspace*{-0.5cm}\includegraphics[scale=0.8]{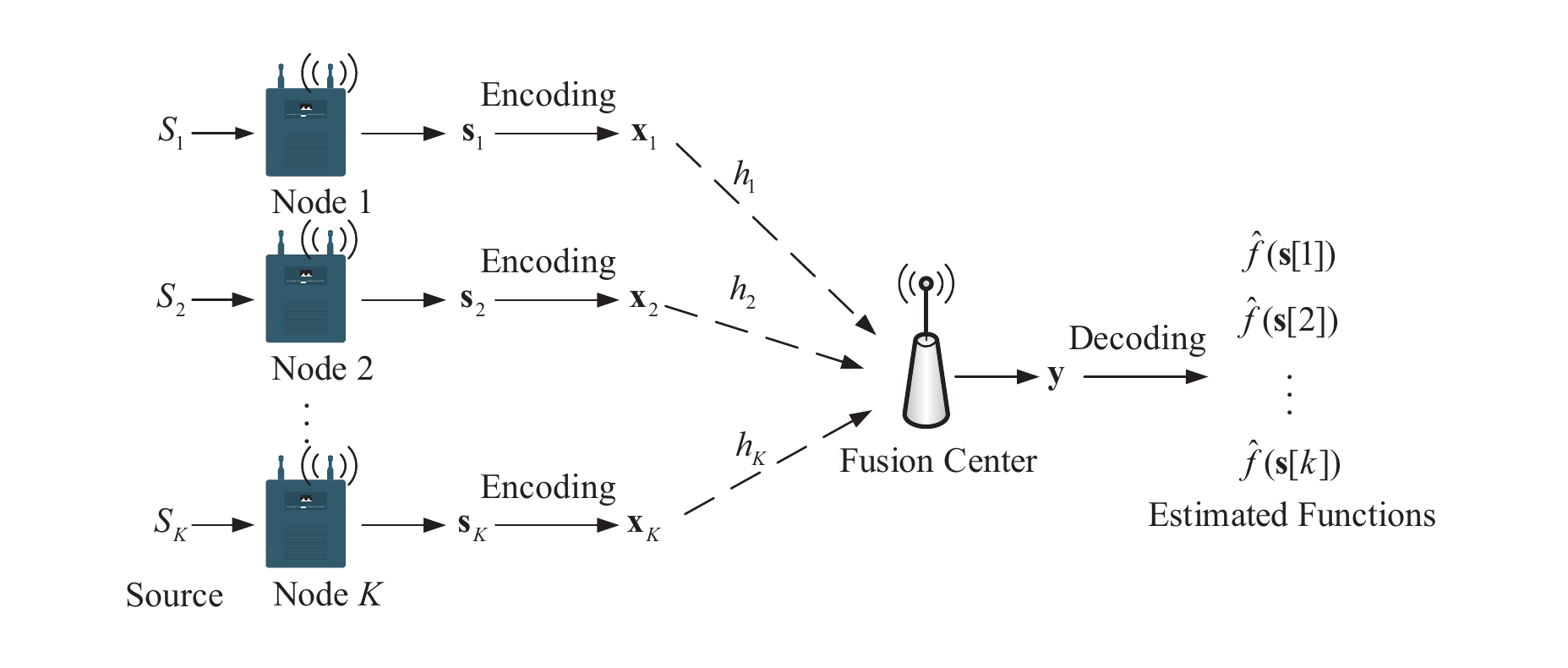}
	\caption{A classical framework of CoMAC.}
	\label{fig:system}
\end{figure}

\subsection{CoMAC}

A classical framework of CoMAC is depicted in Fig.~\ref{fig:system}. In this framework, the fusion center is designed to compute a desired function from $ K $ nodes. Each node obtains data from a corresponding random source for $ T_d $ times, then giving a data vector of length $ T_d $. 

\begin{definition}[Data Matrix]\label{Sources}\label{Discrete sample vector}
	Assuming that $ S_i $ is a random source from which the $ i $-th node gets data, let $ \bm{\mathrm{s_r}}=\left[ S_1, S_2, \cdots, S_K\right]  $ be a random vector associated with a joint probability mass function $ p_{\bm{\mathrm{s_r}}}(\cdot) $. A data matrix $ \bm{\mathrm{S}}\in\mathbb{C}^{T_d\times K}$ represents the data from $ K $ nodes during $ T_d $ time slots. It is expressed as
	\begin{align}
	\bm{\mathrm{S}}&=\begin{bmatrix}
	s_{1}[1] & \cdots & s_{1}[j] & \cdots & s_{1}[T_d] \\ 
	\vdots &  & \vdots &  & \vdots \\ 
	s_{i}[1] & \cdots & s_{i}[j] & \cdots & s_{i}[T_d] \\ 
	\vdots &  & \vdots &  & \vdots \\ 
	s_{K}[1] & \cdots & s_{K}[j] & \cdots & s_{K}[T_d]
	\end{bmatrix} \nonumber\\
	&=\begin{bmatrix}
	{\bm{\mathrm{s}}[1]}^{\mathrm{T}} & \cdots & {\bm{\mathrm{s}}[j]}^{\mathrm{T}} & \cdots & {\bm{\mathrm{s}}[T_d]}^{\mathrm{T}}
	\end{bmatrix} \nonumber\\
	&=\begin{bmatrix}
	\bm{\mathrm{s}}_1^{\mathrm{T}} & \cdots & \bm{\mathrm{s}}_i^{\mathrm{T}} & \cdots & \bm{\mathrm{s}}_K^{\mathrm{T}}
	\end{bmatrix}^{\mathrm{T}},
	\end{align}
	where $ j\in[1:T_d] $, $ i\in[1:K] $, $ s_{i}[j] $ is the $ j $-th data of the $ i $-th node from the random source $ S_i $, $ \bm{\mathrm{s}}[j]=[s_1[j],\cdots,s_K[j]] $ is the $ j $-th data of all $ K $ nodes and $ \bm{\mathrm{s}}_i=\left[s_i[1],\cdots,s_i[T_d] \right] $ is the data vector of node $ i $. Note that $ s_i[j]$ belongs to $[0:p-1] $, which means it is mapped to a number between 0 and $ p-1 $ through quantization. And $ \bm{\mathrm{s}}[j]$ is independently drawn from $ p_{\bm{\mathrm{s_r}}}(\cdot) $.
\end{definition}

Using Definition \ref{Sources}, the desired function determined by the random source vector $ \bm{\mathrm{s_r}} $ can be expressed as $ f(\bm{\mathrm{s_r}}) $, and its definition is in the following.
\begin{definition}[Desired Function]\label{Desired Function D}
	For all $ j\in[1:T_d] $, every function
	\begin{equation}\label{Desired Function}
	f(s_1[j],s_2[j],\cdots,s_K[j])=f(\bm{\mathrm{s}}[j])
	\end{equation}
	that has to be computed at the fusion center is called a desired function where $\bm{\mathrm{s}}[j]$ is independently drawn from $ p_{\bm{\mathrm{s_r}}}(\cdot) $ (See Definition \ref{Sources}). Every function $ f(\bm{\mathrm{s}}[j]) $ can be seen as a realization of $  f(\bm{\mathrm{s_r}})  $. Thus, it has $ T_d $ functions when each node gets data from each random source for $ T_d $ times. 
\end{definition}

\begin{remark}\label{Typical Functions}
	As studied in \cite{giridhar2005computing,wang2015interactive,goldenbaum2014computation,jeon2014computation}, CoMAC can be designed to compute different types of desired function. There are two typical functions that we focus on. A function $ f(\bm{\mathrm{s}}[j])$ whose values are in the set $ \lbrace \sum_{i=1}^{K}a_{1,i}s_i[j], \cdots, \sum_{i=1}^{K}a_{L_s,i}s_i[j] \rbrace  $ is called the arithmetic sum function where $ a_{l,i} $ is the weighting factor for node $ i $ and $ L_s $ belongs to $ \mathbb{N} $. The arithmetic sum function is a weighted sum function, which includes the mean function for all $ K $ nodes $ f(\bm{\mathrm{s}}[j])=\frac{1}{K}\sum_{i=1}^{K}s_i[j] $ and the function for the active node only $ f(\bm{\mathrm{s}}[j])=\left\lbrace s_1[j], s_2[j], \cdots, s_K[j]\right\rbrace $ as special cases. Apart from this, a function $ f(\bm{\mathrm{s}}[j])$ with  values in the set of $\lbrace \sum_{i=1}^{K}\bm{1}_{s_i[j]=0}, \cdots,  \sum_{i=1}^{K}\bm{1}_{s_i[j]=p}\rbrace  $ is regarded as the type function where $ \bm{1}_{(\cdot)} $ denotes the indicator function. As pointed out in \cite{giridhar2005computing}, any symmetric function such as mean, variance, maximum, minimum and median can be attained from the type function.
\end{remark}

Definition \ref{Desired Function D} shows that the desired function $ f(\bm{\mathrm{s}}[j]) $ is computed by all $ K $ nodes participating in the computation. In some cases, we only need to choose  $ M $ nodes to participate in the computation each time. Then the desired function $ f(\bm{\mathrm{s}}[j]) $ has to be divided into $ B=\frac{K}{M}\in \mathbb{N} $ parts. The function which only uses part of the nodes is called sub-function, and its definition is given as follows.

\begin{definition}[Sub-Function]\label{Partitioning Function}
	Let \begin{equation}\label{key}
	\tau_u=\left\lbrace x\in[1:K]:\left|\tau_u\right|=M \right\rbrace
	\end{equation} 
	denote a set where each element $ x $ is the index from the $ M $ chosen nodes. Suppose that $ \bigcup_{u=1}^{B}\tau_u=[1:K] $ and $ \tau_u\bigcap\tau_v =\emptyset$ for all $ u,v\in[1:B] $, A function $ f\left(\left\lbrace s_i[j]\right\rbrace_{i\in\tau_u} \right) $ is said to be a sub-function if and only if there exists a function $ f_c(\cdot) $ satisfying $ f(\bm{\mathrm{s}}[j])=f_c\left(f\left(\left\lbrace s_i[j]\right\rbrace_{i\in\tau_1}\right), f\left(\left\lbrace s_i[j]\right\rbrace_{i\in\tau_2}\right), \cdots, f\left(\left\lbrace s_i[j]\right\rbrace_{i\in\tau_B}\right) \right)  $.
\end{definition}

Thus, the fusion center reconstructs the desired function based on $ B $ sub-functions, when $ K $ nodes are divided into $ B $ parts.

In order to achieve reliable computations against noise, we apply a block code named sequences of nested lattice codes\cite{nazer2011compute} throughout this paper. Based on the block coding, the definitions of encoding and decoding are given as follows.
\begin{definition}[Encoding \& Decoding]\label{Encoding & Decoding}
	Let $ \bm{\mathrm{s}}_i $ denote the data vector for the $ i $-th node whose length is $ T_d $ (see Definition \ref{Discrete sample vector}). Denote $ \bm{\mathrm{x}}_i=[x_i[1],x_i[2],\cdots,x_i[n]] $ as the length-$ n $ transmitted vector for node $ i $. The received vector whose length is $ n $ is given by $ \bm{\mathrm{y}}=[y[1],y[2],\cdots,y[n]] $ at fusion center. Assuming there is a block code with length $ n $, the encoding and decoding functions can be expressed as follows.
	\begin{itemize}
		\item \emph{Encoding Functions}: the univariate function $ \bm{\mathcal{E}}_i(\cdot) $ which generates $ \bm{\mathrm{x}}_i=\bm{\mathcal{E}}_i(\bm{\mathrm{s}}_i) $ is an encoding function of node $ i $. It means $ \bm{\mathrm{s}}_i $ with length $ T_d $ is mapped to a transmitted vector $ \bm{\mathrm{x}}_i $ with length $ n $ for node $ i $.
		\item \emph{Decoding Functions}: the decoding function $ \bm{\mathcal{D}}_j(\cdot) $ is used to estimate the $ j $-th desired function $  f(\bm{\mathrm{s}}[j])  $, which satisfies $ \hat{f}(\bm{\mathrm{s}}[j])=\bm{\mathcal{D}}_j(\bm{\mathrm{y}}) $. It implies the fusion center obtains $ T_d $ desired functions depending on the whole received vector with length $ n $.
	\end{itemize}
\end{definition}

Considering the block code with length $ n $, the definition of computation rate\cite{Jeon2016Opportunistic,jeon2014computation,goldenbaum2015nomographic,goldenbaum2014computation} can be given as follows.
\begin{definition}[Computation rate]\label{Computation rate}
	The computation rate specifies how many function values can be computed per channel use within a predefined accuracy. It can be written as $ R=\lim\limits_{n\rightarrow \infty}\frac{T_d}{n}H(f(\bm{\mathrm{s_r}})) $ where $ T_d $ is the number of function values (see Definition \ref{Desired Function D}), $ n $ is the length of the block code and $ H(f(\bm{\mathrm{s_r}})) $ is the entropy of $ f(\bm{\mathrm{s_r}})  $. Apart from this, $ R $ is achievable only if there is a length-$ n $ block code in order that the probability $ \Pr\left( \bigcup_{j=1}^{T_d}\left\lbrace \hat{f}(\bm{\mathrm{s}}[j]\neq f(\bm{\mathrm{s}}[j]))\right\rbrace\right)  \rightarrow0 $ as $ n $ increases.
\end{definition}

\subsection{Fading MAC with Wide-band signal}\label{System Model}
Wide-band signal transmission has to face frequency selective fading because the coherence bandwidth of the channel is smaller than the signal bandwidth. OFDM is well-suited to provide robustness against fading. Thus, we describe a CoMAC-OFDM system with $ N $ sub-carriers during $ T_{s} $ OFDM symbols while the length of
the block code is $ n $. Then, the $ m $-th received OFDM symbol at the fusion center in the frequency domain can be given by
\begin{equation}\label{fOFDM}
\bm{Y}[m]=\sum_{i=1}^{K}\bm{V}_i[m]\bm{X}_i[m]\bm{H}_i[m]+\bm{W}[m],
\end{equation}
where $ T_{s}= \lceil  \frac{n}{N} \rceil$, $ \bm{V}_i=\mathrm{diag} \left\lbrace\frac{|h_{i,1}[m]|}{h_{i,1}[m]}\sqrt{P_{i,1}[m]},\cdots, \frac{|h_{i,N}[m]|}{h_{i,N}[m]}\sqrt{P_{i,N}[m]} \right\rbrace$ is the power allocation matrix of node $ i $ where each diagonal element is the power allocated to the $ g $-th sub-carrier, $ \bm{H}_i[m]=\mathrm{diag}\left\lbrace h_{i,1}[m],\cdots h_{i,N}[m]\right\rbrace  $ is a diagonal matrix of which the diagonal element is the frequency response of each sub-carrier for node $ i $, $ \bm{X}_i[m]=\mathrm{diag} \left\lbrace x_{i,1}[m], x_{i,2}[m], \cdots, x_{i,N}[m] \right\rbrace  $ is the transmitted diagonal matrix of node $ i $ at the $ m $-th OFDM symbol, the diagonal element $ x_{i,g}[m] $ in $ \bm{X}_i[m] $ represents an element whose index is $ (m-1)N+g $ in the transmitted vector $ \bm{\mathrm{x}}_i $ (See Definition \ref{Encoding & Decoding}) for node $ i $ and the diagonal element of $ \bm{W}[m] $ can be regarded as identically and independently distributed (i.i.d.) complex Gaussian random noise following $ \mathcal{CN}(0,\frac{1}{N}) $.

According to the above expression, the received signal in the $ g $-th sub-carrier at the $ m $-th OFDM symbol can be expressed as
\begin{equation}\label{sOFDM}
y_{g}[m]=\sum_{i=1}^{K}x_{i,g}[m]v_{i,g}[m]h_{i,g}[m]+w_{i,g}[m],
\end{equation}
where $ g\in[1:N] $, $ m\in[1:T_s] $, $ v_{i,g}[m] $ is the $ g$-th diagonal element from $ \bm{V}_i[m] $, $ h_{i,g}[m] $ from $ \bm{H}_i[m] $ is the frequency response of the $ g $-th sub-carrier of node $ i $ at the $ m $-th OFDM symbol, $ w_{i,g}[m] $ is i.i.d. complex Gaussian random noise following $ \mathcal{CN}(0,\frac{1}{N}) $.

The above equations assume perfect synchronization and perfect removal of inter-carrier interference.

\section{Problem Statement \& Main Results}
\label{Problem Statement and Main Results}
In this section, we first introduce the conventional CoMAC schemes, their features and limitations. Thus, we describe our main results and the improvements by comparing them with conventional CoMAC schemes. For easy presentation in following sections, we define the ordered indexes to describe the indexes of nodes which are sorted by the cosponsoring channel gain. For the $ g $-th sub-carrier, let $ \mathcal{I}_{i}^g\in[1:K] $ be the $ i $-th element in the set of ordered indexes of $ [1:K] $ such that $ |h_{\mathcal{I}_{1}^g,g}|\ge|h_{\mathcal{I}_{2}^g,g}|\ge\cdots\ge|h_{\mathcal{I}_{K}^g,g}|  $, e.g., $ \min_{i \in [1:K]}|h_{i,g}|^2=| h_{\mathcal{I}_{K}^g,g}| ^2 $. Further, $ g $ can be omitted if the number of sub-carriers $ N $ is $ 1 $.
\subsection{Previous Works}
In \cite{jeon2014computation}, the authors provided the computation rate for the arithmetic sum and
type functions (See Remark \ref{Typical Functions}) in fading MAC, i.e., the channel coefficients are i.i.d. and vary independently over time.
\begin{theorem}[Conventional CoMAC Rate]\label{Old}
Based on \cite[Theorem 3]{jeon2014computation}, the average computation rate with average power constraint at each time slot is given as
\begin{align}
R=&\mathsf{E}\left[\mathsf{C}^+\left( \dfrac{1}{K}+\dfrac{\min_{i \in [1:K]}\left[ |h_{i}|^2P\right] }{\mathsf{E}\left[\dfrac{\min_{i \in [1:K]}|h_{i}|^2}{|h|^2} \right]}\right)\right]\nonumber\\
\stackrel{(a)}{=}&\mathsf{E}\left[\mathsf{C}^+\left( \dfrac{1}{K}+\dfrac{\left| h_{\mathcal{I}_{K}}\right|^2P }{\mathsf{E}\left[\dfrac{\left| h_{\mathcal{I}_{K}}\right| ^2}{|h|^2} \right]}\right)\right],\label{OLDINC}
\end{align}
where $ \frac{1}{\mathsf{E}\left[|h_{\mathcal{I}_{K}}|^2/|h|^2 \right] }\ge 1$ is the gain from the average power constraint, each channel response of the $ i $-th node $ h_i $ and $ h $ are i.i.d. random variables and the condition $ (a) $ follows on the ordered indexes.
\end{theorem}

Unfortunately, for wireless networks with a massive number of nodes, the computation rate approaches 0 as $ K $ increases, which is a vanishing computation rate. To resolve the problem, an opportunistic CoMAC system has been discussed in \cite{Jeon2016Opportunistic}. In each time slot, only $ M $ nodes which have the largest $ M $ channel gains participate in the transmission such that the indexes of the $ M $ chosen nodes is in the set $ \left\lbrace\mathcal{I}_{i}\right\rbrace_{i\in[1:M]}  $.
\begin{theorem}[Opportunistic CoMAC Rate]\label{Older}
	As shown in \cite[Theorem 1]{Jeon2016Opportunistic}, for any $ M,B\in\mathbb{N} $ satisfying $ MB=K $, the computation rate of the opportunistic CoMAC is given by
\begin{equation}\label{QINC}
R=\dfrac{1}{B}\mathsf{E}\left[ \mathsf{C}^+\left( \dfrac{1}{M}+\dfrac{|h_{\mathcal{I}_{M}}|^2KP}{M\mathsf{E}\left[\dfrac{|h_{\mathcal{I}_{M}}|^2}{|h|^2} \right]}\right)\right],
\end{equation}
where $ \mathcal{I}_{M} $ is the $ M $-th element of the set of ordered indexes $ \left\lbrace\mathcal{I}_{i}\right\rbrace_{i\in[1:K]}  $. 

Theorem \ref{Older} improved the computation rate in fading environments and provided a non-vanishing computation rate as $ K $ increases.
\end{theorem}

However, the wide-band CoMAC has not been discussed and the improvement of computation rate is still limited in \cite{Jeon2016Opportunistic,giridhar2005computing,goldenbaum2014computation,goldenbaum2015nomographic,jeon2014computation,wang2015interactive,appuswamy2014computing,zhan2013linear,ma2012interactive}. Therefore, Our paper focuses on designing a framework of CoMAC for wide-band transmission, which can be robust against both frequency selective fading and the vanishing computation rate. Apart from this, under our sub-function allocation and power allocation, we achieve a higher computation rate and provide a improved non-vanishing computation rate compared with the conventional CoMAC schemes.

\subsection{Main Results}
We employ OFDM design into wide-band CoMAC and divide a desired function into $ B=\frac{K}{M}\in\mathbb{N} $ sub-functions. By allocating these sub-functions into different sub-carriers, the computation rate based on CoMAC-OFDM is presented in Theorem \ref{GINCviaOFDM}.
\begin{theorem}[General CoMAC-OFDM Rate]\label{GINCviaOFDM}
	For any $ M, N\in\mathbb{N} $ satisfying $ M\le K $, the computation rate of wide-band CoMAC-OFDM over fading MAC is given by
	\begin{align}
	R=&\dfrac{M}{KN}\dfrac{1}{T_s}\sum_{m=1}^{T_s}\left[\sum_{g=1}^{N} \mathsf{C}^+\left( \dfrac{N}{M}+N\times\right.\right.\nonumber\\
	&\left.\left.\min_{i \in [1:M]}\left[ |h_{\mathcal{I}_{i}^g[m],g}[m]|^2P_{\mathcal{I}_{i}^g[m],g}[m]\right] \right)\right]\label{OCOMRate}\\
	\stackrel{(a)}{=}&\dfrac{M}{KN}\mathsf{E}\left[\sum_{g=1}^{N} \mathsf{C}^+\left( \dfrac{N}{M}+N\min_{i \in [1:M]}\left[ |h_{\mathcal{I}_{i}^g,g}|^2P_{\mathcal{I}_{i}^g,g}\right] \right)\right],
	\end{align}
	where the condition $ (a) $ follows while $T_s \to \infty$, $ T_s $ is the number of OFDM symbols as shown in Eqs.~\eqref{fOFDM} and \eqref{sOFDM}, $ K $ is the number of nodes for a desired function (See Definition \ref{Desired Function}), $ M $ is the number of the chosen nodes for a sub-function (See Definition \ref{Partitioning Function}), $ N $ is the number of the sub-carriers for each OFDM symbol, $ h_{\mathcal{I}_{i}^g[m],g}[m] $ is the frequency response of the $ g $-th sub-carrier of node $ \mathcal{I}_{i}^g[m] $, $ P_{\mathcal{I}_{i}^g[m],g}[m] $ is the power allocated to the $ \mathcal{I}_{i}^g[m] $-th node in the $ g $-th sub-carrier at the $ m $-th OFDM symbol.
\end{theorem}
\begin{IEEEproof}
	We refer to Section \ref{Sub-Carrier Loading based on OFDM-INC} for proof.
\end{IEEEproof}
\begin{remark}
	Theorem \ref{GINCviaOFDM} is a general computation rate for wide-band CoMAC-OFDM, and the achievable rate for any power allocation method can be derived through Theorem \ref{GINCviaOFDM}. It also generalizes the results for the conventional CoMAC schemes in Theorems \ref{Old} and \ref{Older} by setting the number of sub-carriers $ N $ to 1.
\end{remark}

A simple computation rate for wide-band CoMAC-OFDM system without sub-function allocation and adaptive power allocation can be easily obtained from Theorem \ref{GINCviaOFDM}.
\begin{corollary}[Direct CoMAC-OFDM Rate]\label{ComOFDMINC}
	For a straightforward wide-band CoMAC-OFDM system, each desired function is assigned to each sub-carrier directly and the node $ i $ has an identical average power constraint in the $ g $-th sub-carrier, i.e., $ \mathsf{E}\left[\left|x_{i,g}[m] \right|^2  \right]\le \frac{P}{N}  $ for all $ i\in[1:K]$ and $g\in[1:N] $. The computation rate can be given as
	\begin{equation}\label{RateComOFDMINC}
	R_{C1}=\mathsf{E}\left[\mathsf{C}^+\left( \dfrac{N}{K}+\dfrac{|h_{\mathcal{I}_{K}}|^2P}{\mathsf{E}\left[\dfrac{|h_{\mathcal{I}_{K}}|^2}{|h|^2} \right]}\right)\right].
	\end{equation}
\end{corollary}
\begin{IEEEproof}
	We refer to Section \ref{Average Power Control} for proof.
\end{IEEEproof}

By setting $ N=1 $ in Corollary \ref{ComOFDMINC}, the computation rate is the same as Theorem \ref{Old}. Comparing \eqref{OLDINC} and \eqref{RateComOFDMINC}, the computation rate for wide-band CoMAC-OFDM can be improved by increasing the number of sub-carriers $ N $. However, even though $ N $ can keep increasing, it still has upper limit due to electronic devices. Thus, $ N $ cannot be infinity in practice. The computation rates in Eq.~\eqref{OLDINC} and Eq.~\eqref{RateComOFDMINC} with a fixed $ N $ will become vanishing as the number of nodes $ K $ increases, and converge to 0 in fading MAC.

In order to deal with this, we propose sub-function allocation to avoid the vanishing computation rate even if the number of nodes $ K $ in the network increases. Based on Theorem \ref{GINCviaOFDM}, the computation rate of CoMAC-OFDM with sub-function allocation can be expressed as follows.
\begin{corollary}[CoMAC-OFDM Rate with Sub-Function Allocation]\label{ESubPowerINC}
	Suppose that a desired function is divided into $ B $ sub-functions, and each sub-function is allocated to a sub-carrier where the channel gains of the $ M $ nodes are the largest. The computation rate of wide-band CoMAC-OFDM with sub-function allocation and average power constraint can be given as
	\begin{equation}\label{key}
		R_{C2}=\dfrac{M}{K}\mathsf{E}\left[ \mathsf{C}^+\left( \dfrac{N}{M}+\dfrac{|h_{\mathcal{I}_{M}}|^2KP}{M\mathsf{E}\left[\dfrac{|h_{\mathcal{I}_{M}}|^2}{|h|^2} \right]}\right)\right].
	\end{equation}
\end{corollary}
\begin{IEEEproof}
	We refer to Section \ref{Average Power Control} for proof.
\end{IEEEproof}

By setting $ N=1, K=MB $ in Corollary \ref{ESubPowerINC}, the computation rate is the same as that in Theorem \ref{Older}. Compared with Eq.~\eqref{QINC}, it shows that the proposed scheme can not only provide a non-vanishing computation rate but also improve the computation rate when the number of sub-carriers $ N $ increases.

In addition to average power constraint, OFDM can achieve a higher rate with adaptive power allocation for each node. The instantaneous computation rate for a OFDM symbol is obtained as follows.
\begin{corollary}[CoMAC-OFDM Rate with Sub-Function Allocation and Optimal Power Allocation]\label{TPowerINC}
	The optimal computation rate with sub-function allocation at the $ m $-th OFDM symbol is given as
	\begin{equation}\label{key}
	R_{C3}[m]=\dfrac{M}{KN}\sum_{g=1}^{N} \mathsf{C}^+\left( \dfrac{N}{M}+N\eta^*_g[m]\right),
	\end{equation}
	where $ \eta_g^{*}[m] $ is the optimal level for the  $ g $-th sub-carrier at the $ m $-th OFDM symbol, which will be further explained in Section \ref{Improved OFDM} .
\end{corollary}
\begin{IEEEproof}
	We refer to Section \ref{Improved OFDM solution} for the proof.
\end{IEEEproof}

Depending on the proposed sub-function allocation, we consider an optimization problem to achieve a higher computation rate in OFDM with total power constraint for each node. Based on the solution to the optimization problem, we propose a sponge-squeezing algorithm perform the optimal power allocation method.

\section{Sub-Function Allocation for CoMAC-OFDM}\label{Sub-Carrier Loading based on OFDM-INC}
In this section, we explain the framework of CoMAC-OFDM, which is designed to transmit wide-band signals and improve the computation rate. In Section \ref{Sub-Carrier Loading Scheme}, we present sub-function allocation and expound some technical issues that need to be addressed. Then, we resolve these issues and provide the computation rate in Theorem \ref{GINCviaOFDM} from Section \ref{Section Computation Rate for Partitioning Functions} to Section \ref{Combination of Partitioning Functions}.

\subsection{Sub-Function Allocation}\label{Sub-Carrier Loading Scheme}
\begin{figure}
	\centering
	\subfloat[Function division]{
		\includegraphics[scale=0.8]{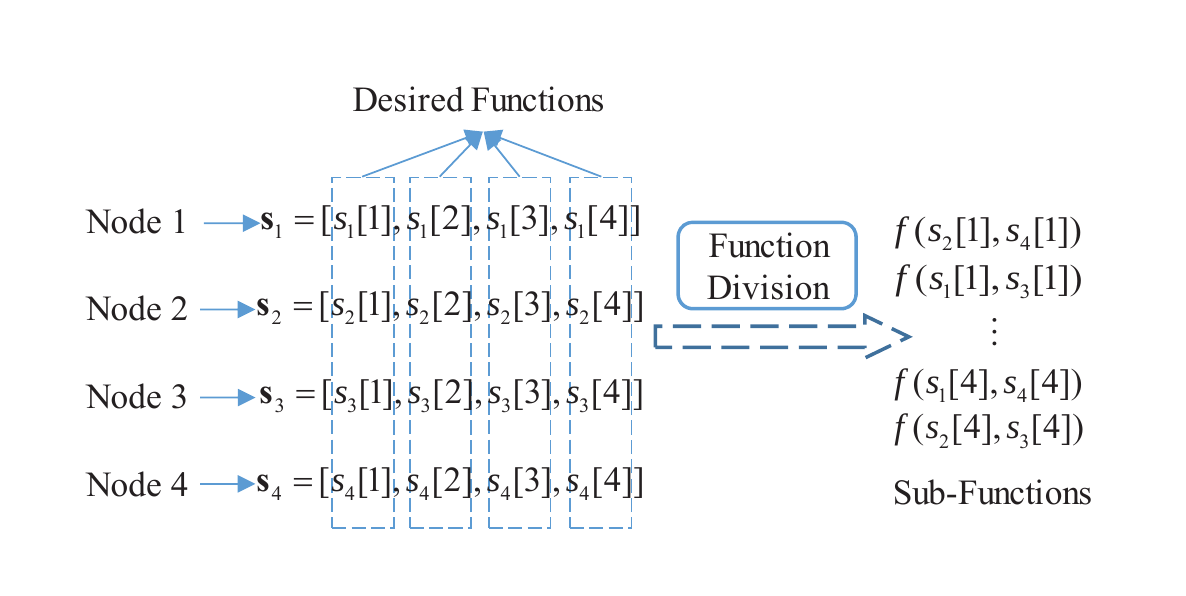}
		\label{fig:ofdm-fd}
	}
	\vfil
	\subfloat[Function allocation]{
		\includegraphics[scale=0.8]{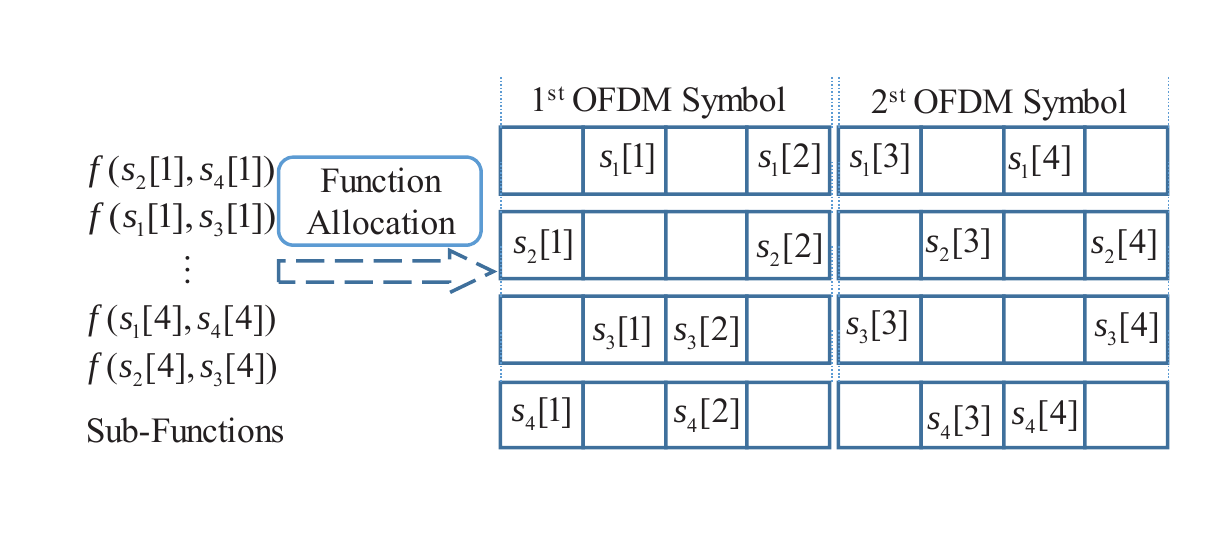}
		\label{fig:ofdm-fa}
	}
	\vfil
	\subfloat[Function reconstruction]{
		\includegraphics[scale=0.8]{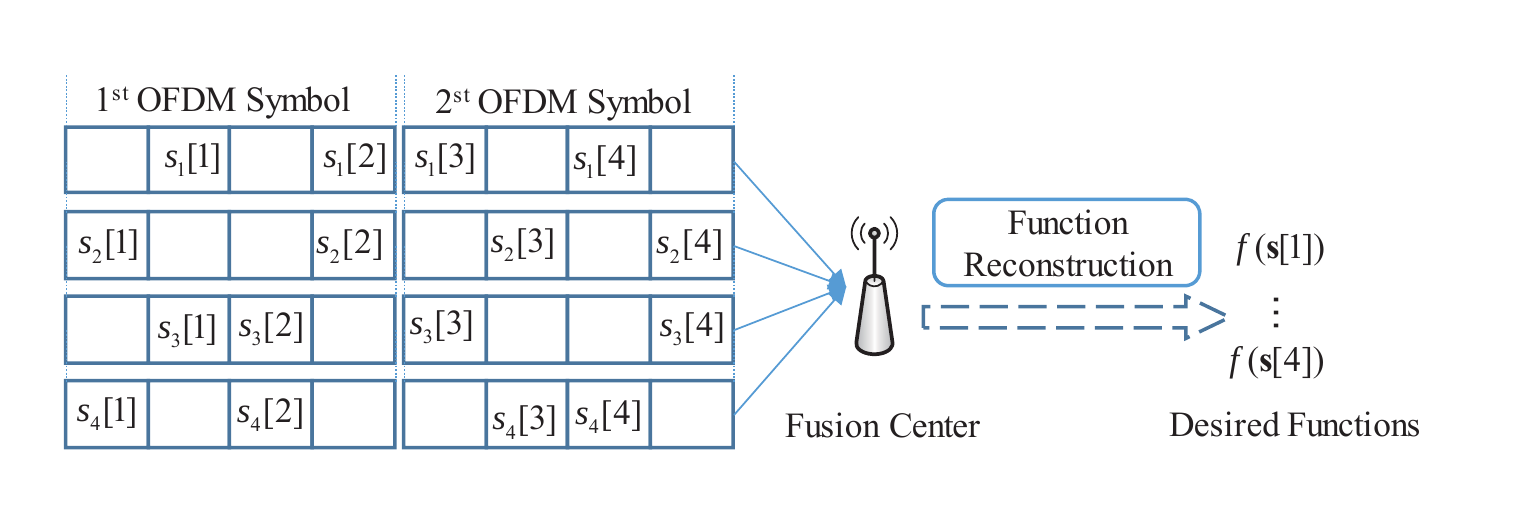}
		\label{fig:ofdm-fc}
	}
	\caption{Implementation of sub-function allocation through division, allocation and reconstruction of function in CoMAC-OFDM.}
	\label{fig:ofdm}
\end{figure}

As shown in Fig.~\ref{fig:ofdm}, sub-function allocation includes three main parts, function division, function allocation and function reconstruction. We provide a simplified description on sub-function allocation based on CoMAC-OFDM system as follows.
\begin{itemize}
	\item \emph{Function Division.} In Fig.~\ref{fig:ofdm-fd}, Node $ i $ gets data from the corresponding random source and obtains the data vector $ \bm{\mathrm{s}}_i $. Then, $ 4 $ desired functions is divided into $ 8 $ sub-functions, which means only $ 2 $ chosen nodes participate in the computation for a sub-function.
	\item \emph{Function Allocation.} Each sub-function is only allocated to the sub-carrier where the channel gains of the chosen nodes are the largest. For a sub-function $ f(s_1[4],s_4[4]) $ in Fig.~\ref{fig:ofdm-fa}, the chosen nodes are Node 1 and Node 4. When the channel gains of Nodes 1 and 4 are the largest ones in the $ 3 $-th sub-carrier at the $ 2 $-th OFDM symbol, e.g., $ \left|h_{1}\right|\ge\left|h_{4}\right|\ge\left|h_{2}\right|\ge\left|h_{3}\right|$, the sub-function $ f(s_1[4],s_4[4]) $ should be assigned to this sub-carrier. After function allocation, all the nodes transmit the corresponding OFDM symbols over the air.
	\item \emph{Function Reconstruction.} At the fusion center in Fig.~\ref{fig:ofdm-fc}, two OFDM symbols are received, and each sub-function can be obtained in each sub-carrier. Using the $ 8 $ received sub-functions, $ 4 $ desired functions are reconstructed by using the relationship between the sub-functions and the desired functions.
\end{itemize}

However, the above description of sub-function allocation is ideal, and the following practical issues should be addressed.
\begin{enumerate}[i]
	\item The order of the channel gains at each sub-carrier is random and depends on the channel realizations in practice. This means that it is important to establish the rules for sub-function allocation.\label{I1}
	\item The above description ignores the encoding and decoding of nested lattice code. It needs employ sequences of nested lattice codes into sub-function allocation.\label{I2}
\end{enumerate}

Further discussion on solving these issues is in Section \ref{Combination of Partitioning Functions}.

\subsection{Computation Rate for Sub-Functions}\label{Section Computation Rate for Partitioning Functions}
In the sub-function allocation shown in Section \ref{Sub-Carrier Loading Scheme}, the fusion center computes one sub-function at a sub-carrier. Each sub-function is only allocated into the sub-carrier where the channel gains of the chosen $ M $ nodes are the largest $ M $ ones. The indexes of those chosen nodes are in a set $ \left\lbrace \mathcal{I}_{i}^g[m]\right\rbrace _{i=1}^M $ for the $ g $-th sub-carrier at the $ m $-th OFDM symbol. For the sake of fairness, we assume the bandwidth of CoMAC-OFDM with N sub-carriers is the same as the bandwidth of the conventional CoMAC system which uses single carrier for transmission. It means the bandwidth of the sub-carrier is $ \frac{1}{N} $ of the bandwidth of the single carrier. Then, the computation rate of a sub-function at the $ m $-th OFDM symbol can be given as follows.
\begin{lemma}\label{Computation Rate for Partitioning Functions}
	When the length of block $ n\to\infty $, the instantaneous computation rate of single sub-function of the $ g$-th sub-carrier at the $ m $-th OFDM symbol with additive white Gaussian noise and a variance of $ \frac{1}{N} $ is given as
	\begin{equation}
	\begin{split}
	R_{\rho,g}[m]=&\frac{1}{N}\mathsf{C}^+\left( \dfrac{N}{M}+N\times\right.\\
	&\left.\min_{i \in [1:M]}\left[ |h_{\mathcal{I}_{i}^g[m],g}[m]|^2P_{\mathcal{I}_{i}^g[m],g}[m]\right] \right),
	\end{split}	
	\end{equation}
	where $ h_{\mathcal{I}_{i}^g[m],g}[m] $ is the frequency response of the $ g $-th sub-carrier for node $ \mathcal{I}_{i}^g[m] $ (See Eq.~\eqref{sOFDM}),  $P_{\mathcal{I}_{i}^g[m],g}[m] $ is the allocated power of the $ \mathcal{I}_{i}^g[m] $-th node in the $ g $-th sub-carrier at the $ m $-th OFDM symbol.
\end{lemma}
\begin{IEEEproof}
	\cite[Theorem 4, Eq. (85)]{nazer2011compute} shows that the computation rate with a noise variance of $ \sigma_Z^2 $ is given as
	\begin{equation}\label{key}
	R=\mathsf{C}^+\left(\dfrac{1}{\sigma_Z^2} \left( \left\| \bm{a}\right\| ^2-\dfrac{P\left|\bm{h}^H\bm{a} \right|^2 }{1+P\left\| \bm{h}\right\|^2}\right)^{-1}  \right), 
	\end{equation}
	where $ \bm{a} $ is the equation coefficient vector, $ \bm{h} $ is channel coefficient vector and $ P $ is the power of each transmitted symbol.
	
	By setting $ \sigma_Z^2=\frac{1}{N}, \bm{a}=\bm{h} $ and using \cite[Theorems 4 and 11]{nazer2011compute}, the computation rate can be written as
	\begin{equation}\label{partNosie}
	R=\mathsf{C}^+\left( \dfrac{N}{M}+NP\right).
	\end{equation}
	Then by combining Eq.~\eqref{partNosie} with \cite[Theorem 3]{jeon2014computation}, the instantaneous computation rate in fading MAC with additive white Gaussian noise whose variance is $ \frac{1}{N} $ can be expressed as
	\begin{equation}\label{single carrier}
	R[t]=\mathsf{C}^+\left( \dfrac{N}{M}+N
	\min_{i \in [1:M]}\left[ |h_{\mathcal{I}_{i}[t]}[t]|^2P_{\mathcal{I}_{i}[t]}[t]\right] \right)
	\end{equation}
	at the $ t $-th time slot. Furthermore, the symbol of a sub-carrier in OFDM lasts $ N $ time slots while the symbol in the single carrier lasts 1 time slot. According Definition \ref{Computation rate}, the number of the function values in a sub-carrier is $ \frac{1}{N} $ of the number of the function values in single carrier, which means the computation rate of a sub-carrier is $ \frac{1}{N} $ of the computation rate in Eq.~\eqref{single carrier}. In conclusion, Lemma \ref{Computation Rate for Partitioning Functions} has been proved.
\end{IEEEproof}

\subsection{Reconstruction of Sub-Functions}\label{Combination of Partitioning Functions}
Only $ M $ nodes are chosen from all $ K $ nodes for a sub-function, which means there need be $ B=\frac{K}{M} $ sub-functions to reconstruct a desired function. $ \tau $ denotes a set where the elements are the indexes of the $ M $ chosen nodes for a sub-function $ f\left(\left\lbrace s_i[j]\right\rbrace_{i\in\tau} \right) $ (See Definition \ref{Partitioning Function}). Then, the set
\begin{equation}\label{key}
\mathcal{S}=\left\lbrace \tau\subseteq[1:K]:\left|\tau \right|=M \right\rbrace 
\end{equation}
includes all the possible sub-functions\footnote{For easy presentation, we use the element $ \tau\in\mathcal{S} $ stands for the sub-function $ f\left(\left\lbrace s_i[j]\right\rbrace_{i\in\tau} \right) $ which is computed by these nodes in $ \tau $.}, and the cardinality of $ \mathcal{S} $ is $ \left| \mathcal{S}\right|=\left(_M^K\right) $. According to Definition \ref{Partitioning Function}, the sub-functions can be reconstructed into the desired function if and only if $\bigcup_{u=1}^{B}\tau_u=[1:K] $ and $ \tau_u\bigcap\tau_v =\emptyset$ for all $ u,v\in[1:B] $ and $\tau_u,\tau_v\in\mathcal{S}  $. We define \begin{equation}\label{SetQ}
\mathcal{Q}=\left\lbrace\rho=\left\lbrace\tau_1,\tau_2,\cdots,\tau_B \right\rbrace:\bigcup_{u=1}^{B}\tau_u=[1:K], \tau_u\in\mathcal{S} \right\rbrace, 
\end{equation}
which contains all possible combinations that can be used to reconstruct the desired function, and $  \left| \mathcal{Q}\right|=\prod_{l=0}^{B-1}\left(_M^{K-Ml}\right) $. 

Under our design, the length-$ n $ block code is transmitted through $ T_s $ OFDM symbols. In order to simplify the derivation, the number of OFDM symbols $ T_s=\frac{n}{N} $ instead of $ \lceil\frac{n}{N}\rceil $ in Eq.~\eqref{fOFDM}, which means the number of all the sub-carriers during $ T_s $ OFDM symbols is $ n $. Based on sub-function allocation, all the sub-carriers should be assigned to every sub-function. The set $ \mathcal{M}_{\tau}$ for all $ \tau \in \mathcal{S}$ includes those sub-carriers are assigned to sub-function $ \tau $. And let a set $ \mathcal{M}_{\rho}^{\tau} $ include the sub-carriers that are from $ \mathcal{M}_{\tau} $ and allocated in the combination $ \rho\in\mathcal{Q} $ that contains the corresponding sub-function $ \tau $.

 As mentioned in Issue \ref{I1}, Section \ref{Sub-Carrier Loading Scheme}, the order of the indexes for $ K $ nodes in each sub-carrier is random depending on channel realizations in practice. It means the number of the sub-carriers in $ \mathcal{M}_{\tau} $ is random. Thus, we use the following lemma to characterize the the minimum deterministic values of $ \left|\mathcal{M}_{\tau} \right|  $ and $ \left| \mathcal{M}_{\rho}^{\tau}\right|  $.
\begin{lemma}\label{UniforPr}
	When $ n $ is large, for all $ \rho\in\mathcal{Q} $ and $ \tau\in{\rho} $, each set $ \mathcal{M}_{\rho}^{\tau} $ contains
	\begin{equation}\label{key}
	\left|\mathcal{M}_{\rho}^{\tau} \right|=\dfrac{\left|\mathcal{M}_{\tau} \right|}{B\frac{\left|\mathcal{Q} \right|}{\left|\mathcal{S} \right|}}=\dfrac{n}{B\left| \mathcal{Q}\right| }
	\end{equation}
	sub-carriers.
\end{lemma}
\begin{IEEEproof}
	The probability, $ \Pr(\left\lbrace \mathcal{I}_{i}^g[m]\right\rbrace _{i=1}^M=\tau) $, for all $ \tau\in \mathcal{S} $ is $ \frac{1}{\left| \mathcal{S}\right|} $ since channel gains are i.i.d. in the $ g $-th sub-carrier at the $ m $-th OFDM symbol. Thus, $ \frac{\left|\mathcal{M}_{\tau} \right|}{n}=\frac{1}{\left| \mathcal{S}\right|} $ holds depending on \cite[Lemma 2.12]{csiszar2011information}. It shows $ \left|\mathcal{M}_{\tau} \right|=\frac{n}{\left| \mathcal{S}\right|} $ for all $ \tau\in\mathcal{S} $ as n increases. Besides, in the set $ \mathcal{Q} $, there are $B\frac{\left|\mathcal{Q} \right|}{\left|\mathcal{S} \right|}  $ combinations which include $ \tau $ as an element for all $ \tau\in\mathcal{S} $. Therefore, each combination $ \rho$ including sub-function $ \tau $ is assigned $ \frac{\left|\mathcal{M}_{\tau} \right|}{B\frac{\left|\mathcal{Q} \right|}{\left|\mathcal{S} \right|}} $ sub-carriers.
\end{IEEEproof}

 According to the definition of $ \mathcal{M}_{\rho}^{\tau} $, we find that there are $ \left|\mathcal{M}_{\rho}^{\tau} \right| $ sub-carriers which are used to transmit sub-function $ \tau $ in the combination $ \rho $ over $ T_s $ OFDM symbols. It also implies that the length of the transmitted signal vector of each node for sub-function $ \tau\in\rho $ is $ \left|\mathcal{M}_{\rho}^{\tau}\right| $. To simplify the derivation, the transmitted signal vector is sent centrally in $ T_{\rho}=\left\lceil\frac{\left|\mathcal{M}_{\rho}^{\tau}\right|}{N}\right\rceil $ OFDM symbols in wide-band CoMAC-OFDM. From Eq.~\eqref{sOFDM}, the received vector for sub-function $ \tau $ based on the combination $ \rho $ at the $ m $-th OFDM symbol in the frequency domain can be given as
\begin{equation}\label{sub-OFDM}
\bm{Y}_{\rho}^{\tau}[m]=\sum_{i\in\tau}\bm{V}_{i,\rho}^{\tau}[m]\bm{X}_{i,\rho}^{\tau}[m]\bm{H}_{i,\rho}^{\tau}[m]+\bm{W}_{\rho}^{\tau}[m].
\end{equation}
Then from Eq.~\eqref{sOFDM}, the received signal in $ g $-th sub-carrier can be given as
\begin{align}
y_{\rho,g}^{\tau}[m]&=\sum_{i\in\tau}x_{i,\rho,g}^{\tau}[m]v_{i,\rho,g}^{\tau}[m]h_{i,\rho,g}^{\tau}[m]+w_{i,\rho,g}^{\tau}[m]\nonumber\\
&=\sum_{i\in\tau}x_{i,\rho,g}^{\tau}[m]h_{i,\rho,g}^{'\tau}[m]+w_{i,\rho,g}^{\tau}[m]\label{sub-sub-OFDM},
\end{align}
where $ h_{i,\rho,g}^{'\tau}[m] $ is $ |h_{i,\rho,g}^{\tau}[m]|\sqrt{P_{i,\rho,g}^{\tau}[m]} $. From the conclusion of Lemma \ref{Computation Rate for Partitioning Functions} and Eqs.~\eqref{sub-OFDM}\eqref{sub-sub-OFDM} , the average rate for computing a sub-function $ f\left(\left\lbrace s_i[j]\right\rbrace_{i\in\tau} \right) $ during $ T_{\rho} $ OFDM symbols is given as 
\begin{align}
R_{\rho}&=\dfrac{1}{T_{\rho}}\sum_{m=1}^{T_{\rho}}\dfrac{1}{N}\sum_{g=1}^{N}\mathsf{C}^+\left( \dfrac{N}{M}+N
\min_{i \in \tau}\left[ |h_{i,g}[m]|^2P_{i,g}[m]\right] \right)\nonumber\\
&\stackrel{(a)}{=}\dfrac{1}{T_{\rho}}\sum_{m=1}^{T_{\rho}}\dfrac{1}{N}\sum_{g=1}^{N}R_{\rho,g}[m]\label{Ex rate for part function via OFDM},
\end{align}
where the condition $ (a) $ follows because $ \left\lbrace \mathcal{I}_{i}^g[m]\right\rbrace _{i=1}^M=\tau $.
 
Note that we employ nested lattice code in the proposed system and the length of the transmitted signal vector of each node in $ \tau\in\rho $ is $ \left|\mathcal{M}_{\rho}^{\tau}\right| $. In order to have this length for the transmitted signal vector, the corresponding length of data vector should be $ U_{\rho}=\frac{R_\rho\left|\mathcal{M}_{\rho}^{\tau}\right|}{H(f(\bm{\mathrm{s_r}}))} $ depending on Definition \ref{Computation rate}, which addresses Issue \ref{I2}, Section \ref{Sub-Carrier Loading Scheme}. Thus, a desired function $ f(\bm{\mathrm{s}}[j]) $ reconstructed by combination $ \rho $ can be described as $ f_c(\left\lbrace f(\left\lbrace s_i[j] \right\rbrace_{i\in\tau} ) \right\rbrace_{\tau\in\rho} ) $ (See Definition \ref{Partitioning Function}), and the number of desired function values reconstructed by combination $ \rho $ is $ U_{\rho} $. Then, the number of desired function values for all $ \rho\in\mathcal{Q} $ is
\begin{align}
T_d&=\sum_{\rho\in\mathcal{Q}}U_{\rho}\nonumber\\
&=\sum_{\rho\in\mathcal{Q}}\dfrac{R_\rho\left|\mathcal{M}_{\rho}^{\tau}\right|}{H(f(\bm{\mathrm{s_r}}))}\label{rate for part function via OFDM}
\end{align}
during $ T_s $ OFDM symbols. As a result, the computation rate based on Definition \ref{Computation rate} for computing the desired function via CoMAC-OFDM
\begin{align}
R&=\lim\limits_{n\rightarrow \infty}\dfrac{T_d}{n}H(f(\bm{\mathrm{s_r}}))\nonumber\\
&\stackrel{(a)}{=}\lim\limits_{n\rightarrow \infty}\dfrac{\displaystyle\sum_{\rho\in\mathcal{Q}}\dfrac{R_\rho\left|\mathcal{M}_{\rho}^{\tau}\right|}{H(f(\bm{\mathrm{s_r}}))}}{n}H(f(\bm{\mathrm{s_r}}))\nonumber\\
&\stackrel{(b)}{=}\lim\limits_{n\rightarrow \infty}\dfrac{1}{B}\dfrac{1}{T_{\rho}\left| \rho\right|\left| \mathcal{Q}\right| }\sum_{\rho\in\mathcal{Q}}\sum_{\tau\in\rho}\sum_{m=1}^{T_{\rho}}\dfrac{1}{N}\sum_{g=1}^{N}R_{\rho,g}[m]\nonumber\\
&\stackrel{(c)}{=}\lim\limits_{n\rightarrow \infty}\dfrac{1}{BN}\dfrac{1}{T_{s} }\sum_{m=1}^{T_{s}}\sum_{g=1}^{N}R_{\rho,g}[m]\nonumber\\
&\stackrel{(d)}{=}\dfrac{1}{BN}\mathsf{E}\left[\sum_{g=1}^{N}R_{\rho,g}\right]\nonumber
\end{align}
is achievable as $ n $ increases where the condition $ (a) $ follows because of Lemma \ref{UniforPr} and Eq.~\eqref{rate for part function via OFDM},  the condition $ (b) $ follows from Eq.~\eqref{Ex rate for part function via OFDM}, the cardinality of $ \rho $ is $ \left| \rho\right|=B  $ from Eq.~\eqref{SetQ}, the condition $ (c) $ follows while $ T_s=T_{\rho}B\left| \mathcal{Q}\right| $ and the condition $ (d) $ follows due to the increase of $ n $. Hence, Theorem \ref{GINCviaOFDM} is proved.
 
\section{Power Allocation for Sub-Function Allocation}
\label{Power Allocation Methods for Sub-Carrier Loading}
In this section, we propose two power allocation methods to implement CoMAC-OFDM with sub-function allocation. For average power allocation, the achievable rate is give in Section \ref{Average Power Control}. We further consider an optimization problem and propose a sponge-squeezing algorithm which provides a simple and concise interpretation of the necessary optimality conditions based on the solution in Section \ref{Improved OFDM}.

\subsection{Average Power Allocation}\label{Average Power Control}
Based on sub-function allocation discussed in Section \ref{Sub-Carrier Loading Scheme}, we proposed a power allocation method with average power constraint.

Because only $ M $ nodes participate in computing a sub-function in each sub-carrier depending on Theorem \ref{GINCviaOFDM}, let $ P_{i,g}[m] $ represent the transmitted power in $ g $-th sub-carrier at the $ m $-th OFDM symbol for the node $ i $, which can be given as
\begin{equation}\label{key}
P_{\mathcal{I}_{i}^g[m],g}[m]=\left\lbrace 
\begin{aligned}
&c\dfrac{|h_{\mathcal{I}_{M}^g[m],g}[m]|^2}{|h_{\mathcal{I}_{i}^g[m],g}[m]|^2}&i\in[1:M]\\
&0&i\in[M+1:K]
\end{aligned}\right..
\end{equation}

In order to compare it with the conventional CoMAC which uses the single carrier to transmit signal in a fair way, we assume the bandwidth of OFDM symbols with $ N $ sub-carriers is the same as the bandwidth of the conventional single carrier symbols. Then, a complete OFDM symbol should use $ N $ time slots to be transmitted while the number of transmitted symbols should be $ N $ in the single carrier system. Hence, the average power in the $ g $-th sub-carrier at the $ m $-th OFDM symbol is $ \mathsf{E}[x_{i,g}[m]] =\frac{P}{N} $. From
\begin{align}\label{key}
\mathsf{E}[P_{i,g}[m]]&=\sum_{j=1}^{K}\Pr(i=\mathcal{I}_{j}^g[m])\mathsf{E}[P_{i,g}[m]|i=\mathcal{I}_{j}^g[m]] \nonumber\\
&=\frac{c}{K}\sum_{j=1}^{M}\mathsf{E}\left[\dfrac{|h_{\mathcal{I}_{M}^g[m],g}[m]|^2}{|h_{\mathcal{I}_{i}^g[m],g}[m]|^2} \right] 
\end{align}
with $ \mathsf{E}[x_{i,g}[m]]\le \frac{P}{N} $, we can calculate $ c $ and set
\begin{equation}\label{APowerControl}
P_{\mathcal{I}_{i}^g[m],g}[m]=\left\{
\begin{aligned}
&\dfrac{KP\dfrac{|h_{\mathcal{I}_{M}^g[m],g}[m]|^2}{|h_{\mathcal{I}_{i}^g[m],g}[m]|^2}}{N\displaystyle\sum_{j=1}^{M}\mathsf{E}\left[\dfrac{|h_{\mathcal{I}_{M}^g[m],g}[m]|^2}{|h_{\mathcal{I}_{j}^g[m],g}[m]|^2} \right]}&i \in [1:M] \\
&0&i\in[M+1:K]
\end{aligned}
\right..
\end{equation}

We further put $ P_{\mathcal{I}_{i}^g[m],g}[m] $ from Eq.~\eqref{APowerControl} into Eq.~\eqref{OCOMRate} in Theorem \ref{GINCviaOFDM}, the computation rate for CoMAC-OFDM with sub-carrier loading scheme can be written as
\begin{align}\label{ProofTh2}
R_{C2}&=\dfrac{M}{KN}\mathsf{E}\left[\sum_{g=1}^{N} \mathsf{C}^+\left( \dfrac{N}{M}+ \dfrac{KNP|h_{\mathcal{I}_{M}^g,g}|^2}{N\sum_{j=1}^{M}\mathsf{E}\left[\dfrac{|h_{\mathcal{I}_{M}^g,g}|^2}{|h_{\mathcal{I}_{j}^g,g}|^2} \right]}\right)\right]\nonumber\\
&\stackrel{(a)}{=}\dfrac{M}{K}\mathsf{E}\left[ \mathsf{C}^+\left( \dfrac{N}{M}+\dfrac{|h_{\mathcal{I}_{M}}|^2KP}{M\mathsf{E}\left[\dfrac{|h_{\mathcal{I}_{M}}|^2}{|h|^2} \right]}\right)\right],
\end{align}
where the condition $ (a) $ follows because the channel gains in different sub-carriers are i.i.d. random variables. In conclusion, Eq.~\eqref{ProofTh2} is achievable as $ T_s $ increases (See Theorem \ref{GINCviaOFDM}), which completes the proof of Corollary \ref{ESubPowerINC}. Particularly, Corollary \ref{ComOFDMINC} can be proved by setting $ M=K $, which means each sub-carrier is allocated a desired function. 
\subsection{Optimal Power allocation}\label{Improved OFDM}
Based on the above analysis, we demonstrate the computation rate based on CoMAC-OFDM in Section \ref{Combination of Partitioning Functions}. We further find that the computation rate using on the average power constraint for each node in each sub-carrier is not optimal. Therefore, we consider the optimal power allocation for CoMAC-OFDM, discuss an optimization problem and propose an algorithm called sponge-squeezing based on the solution.

\subsubsection{Problem Formulation}\label{Improved OFDM solution}
Corollary \ref{GINCviaOFDM} shows that the computation rate in fading MAC is the mean of the instantaneous computation rates, which means Eq.~\eqref{OCOMRate} can be rewritten as
\begin{equation}\label{key}
R=\dfrac{1}{T_s}\sum_{m=1}^{T_s}R[m],
\end{equation}
where the instantaneous computation rate at the $ m $-th OFDM symbol is
\begin{equation}\label{instantaneous computation rate}
\begin{split}
R[m]=&\dfrac{M}{KN}\sum_{g=1}^{N} \mathsf{C}^+\left( \dfrac{N}{M}+N\times\right.\\
&\left.\min_{i \in [1:M]}\left[ |h_{\mathcal{I}_{i}^g[m],g}[m]|^2P_{\mathcal{I}_{i}^g[m],g}[m]\right] \right).
\end{split}
\end{equation}
It implies that our power allocation should focus on the current OFDM symbol depending on the channel gains of each sub-carrier. Thus, the optimization problem based on sub-function loading is given as
\begin{align}
\mathop{\mathrm{maximize}}\limits_{P_{\mathcal{I}_{i}^g[m],g}[m]}& \quad R[m] \nonumber \\
s.t.&\quad \sum_{g=1}^{N}P_{i,g}[m]\le P\quad \forall i\in[1:K].
\end{align}
From Eq.~\eqref{instantaneous computation rate}, we find that the problem is difficult to solve, because the $ \min $ function is nested in the $ \log $  function in the objective function and the ordered indexes $ \left\lbrace \mathcal{I}_{i}^g[m]\right\rbrace_{i\in[1:M]}  $ also create difficulty. In order to make it tractable, we introduce a $ K\times N $ sub-function allocation matrix $ \bm{\omega} $, and each element is given as
\begin{equation}\label{key}
\omega_{\mathcal{I}_{i}^g[m],g}=\left\lbrace 
\begin{aligned}
1\quad &i\in[1:M]\\
0\quad &i\in[M+1:K]
\end{aligned}\right.\quad\forall g\in[1:N].
\end{equation}
Furthermore, $ \min_{i \in [1:M]}\left[ |h_{\mathcal{I}_{i}^g[m],g}[m]|^2P_{\mathcal{I}_{i}^g[m],g}[m]\right]  $ in Eq.~\eqref{instantaneous computation rate} implies that the computation rate is only determined by the node which has the minimum product for the $ M $ chosen nodes in the $ g $-th sub-carrier. It means other nodes need not put much power into this sub-carrier except for the node which has the minimum product, and the unused power can be allocated to the other sub-carriers to improve the rate. Thus, we use the level $ \eta_g[m] $ to replace the $ \min $ function by making $ |h_{\mathcal{I}_{i}^g[m],g}[m]|^2P_{\mathcal{I}_{i}^g[m],g}[m]=\eta_g[m] $ for all $ i\in[1:M] $ in the $ g $-th sub-carrier. Then, the optimization objective can be written as
\begin{equation}\label{key}
	R_{C3}[m]=\dfrac{M}{KN}\sum_{g=1}^{N} \mathsf{C}^+\left( \dfrac{N}{M}+N\eta_g[m] \right).
\end{equation}
Therefore, the problem can be expressed as:
\begin{align}\label{OP}
\mathop{\mathrm{maximize}}\limits_{\eta_g[m]}& \quad R_{C3}[m] \nonumber \\
s.t.&\quad \sum_{g=1}^{N}G_{i,g}[m]\eta_g[m]\omega_{i,g}[m]\le P\quad \forall i\in[1:K],
\end{align}
where $ \eta_g[m]\ge0 $ and $ P_{i,g}[m]\ge0 $ are necessary, $ G_{i,g}[m]=1/|h_{i,g}[m]|^2$, $ P_{i,g}[m]= G_{i,g}[m]\eta_g[m] $, $ \omega_{i,g}[m] $ is the element of the sub-function allocation matrix at the $ m $-th OFDM symbol.

Note that the objective function in Eq.~\eqref{OP} is concave since it is a sum of $ \log $ functions, which are concave themselves. Besides, the constraint set is convex as it is composed of linear constraints. Hence, the above optimization problem is a convex optimization problem, and has a unique maximizer. The Lagrangian function
\begin{dmath}\label{key}
\mathcal{L}=\dfrac{M}{KN}\sum_{g=1}^{N} \mathsf{C}^+\left( \dfrac{N}{M}+N\eta_g[m] \right)-\sum_{i=1}^{K}
\mu_i\left(\sum_{g=1}^{N}G_{i,g}[m]\eta_g[m]\omega_{i,g}[m]-P \right) 
\end{dmath}
with the complimentary slackness condition
\begin{equation}\label{key}
\mu_i\left(\sum_{g=1}^{N}G_{i,g}[m]\eta_g[m]\omega_{i,g}[m]-P=0 \right)\quad\forall i\in[1:K]
\end{equation}
is given, where $ \left\lbrace \mu_i\right\rbrace  $ are Lagrange multipliers.

We apply the KKT optimality conditions to the Lagrangian function to obtain the optimal level $ \eta_g^{*}[m] $ for all $ g\in[1:N] $ in terms of the Lagrange
multipliers $ \left\lbrace \mu_i \right\rbrace_{i=1}^K$. Then, the optimal level $ \eta_g^{*}[m]$ is expressed as
\begin{equation}\label{OPeta}
\eta_g^{*}[m]=\max\left\lbrace0, v_g-\frac{1}{M}\right\rbrace,
\end{equation}
where
\begin{equation}\label{Vg}
v_g=\dfrac{M}{KN\sum_{i=1}^{K}\mu_iG_{i,g}[m]\omega_{i,g}[m]},
\end{equation}
while  $ \eta_g^{*}[m] $ satisfies 
\begin{equation}\label{Constrain}
\max_{i=1}^{K}\left[ \sum_{g=1}^{N}G_{i,g}[m]\eta_g[m]\omega_{i,g}[m] \right]-P=0.
\end{equation} 
Note that the optimal level $ \eta_g^{*}[m] $ is determined by the dot product of the Lagrange multiplier vector and the channel gain vector, which is different with the conventional solution determined by the single Lagrange multiplier or the sum of the Lagrange multiplier vector.

Eqs.~\eqref{OPeta} and \eqref{Constrain} also imply that the optimal power in each sub-carrier for each node is
\begin{equation}\label{OPpower}
P^{*}_{i,g}[m]=G_{i,g}[m]\eta_g^{*}[m]\omega_{i,g}[m].
\end{equation}
In conclusion, Corollary \ref{TPowerINC} holds.

\begin{figure}
	\centering
	\subfloat[Initialization phase]{
	\hspace*{-0.6cm}\includegraphics[scale=0.6]{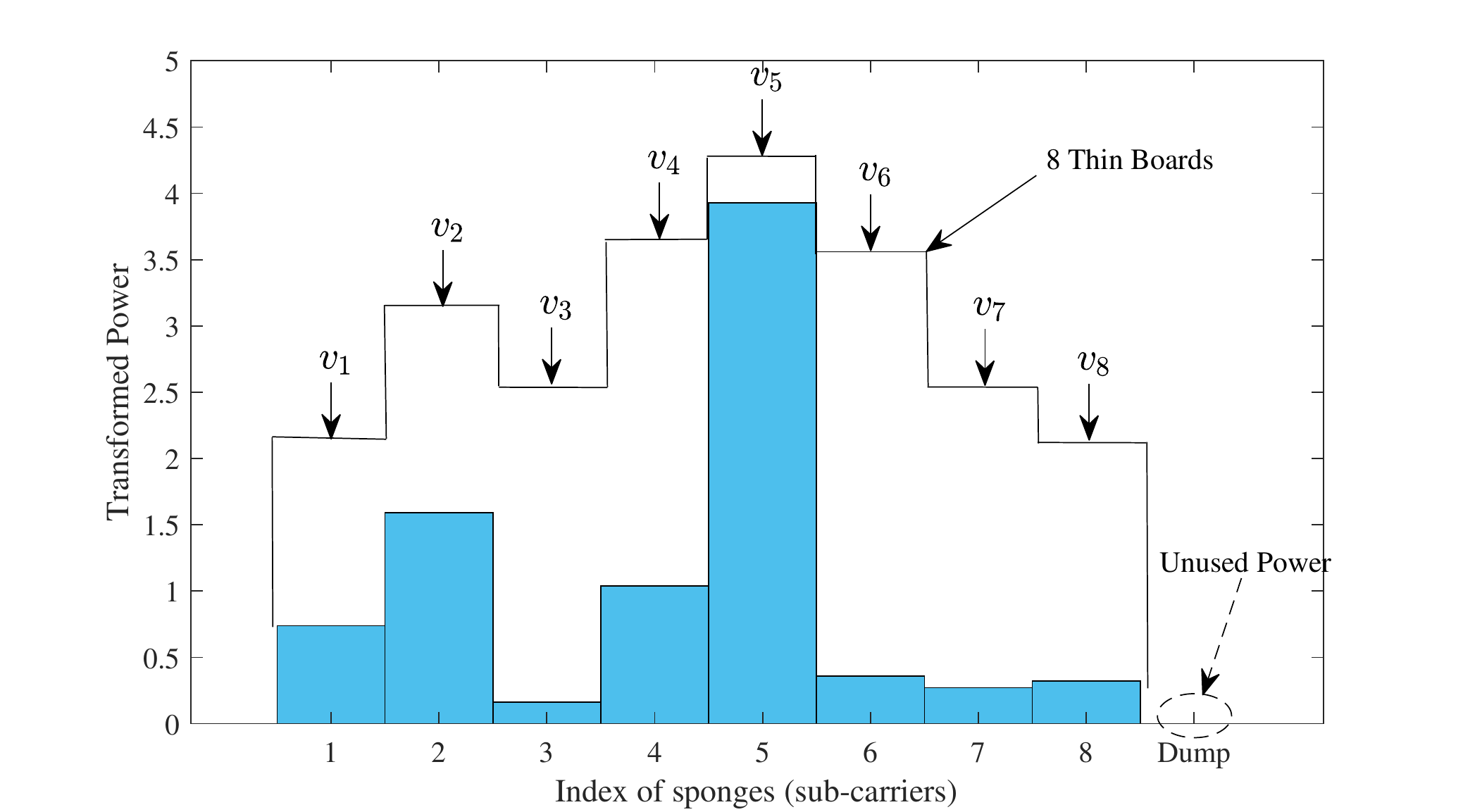}
	\label{fig:ss1}
	}
	\hfil
	\subfloat[Squeezing phase]{
		\hspace*{-0.6cm}\includegraphics[scale=0.6]{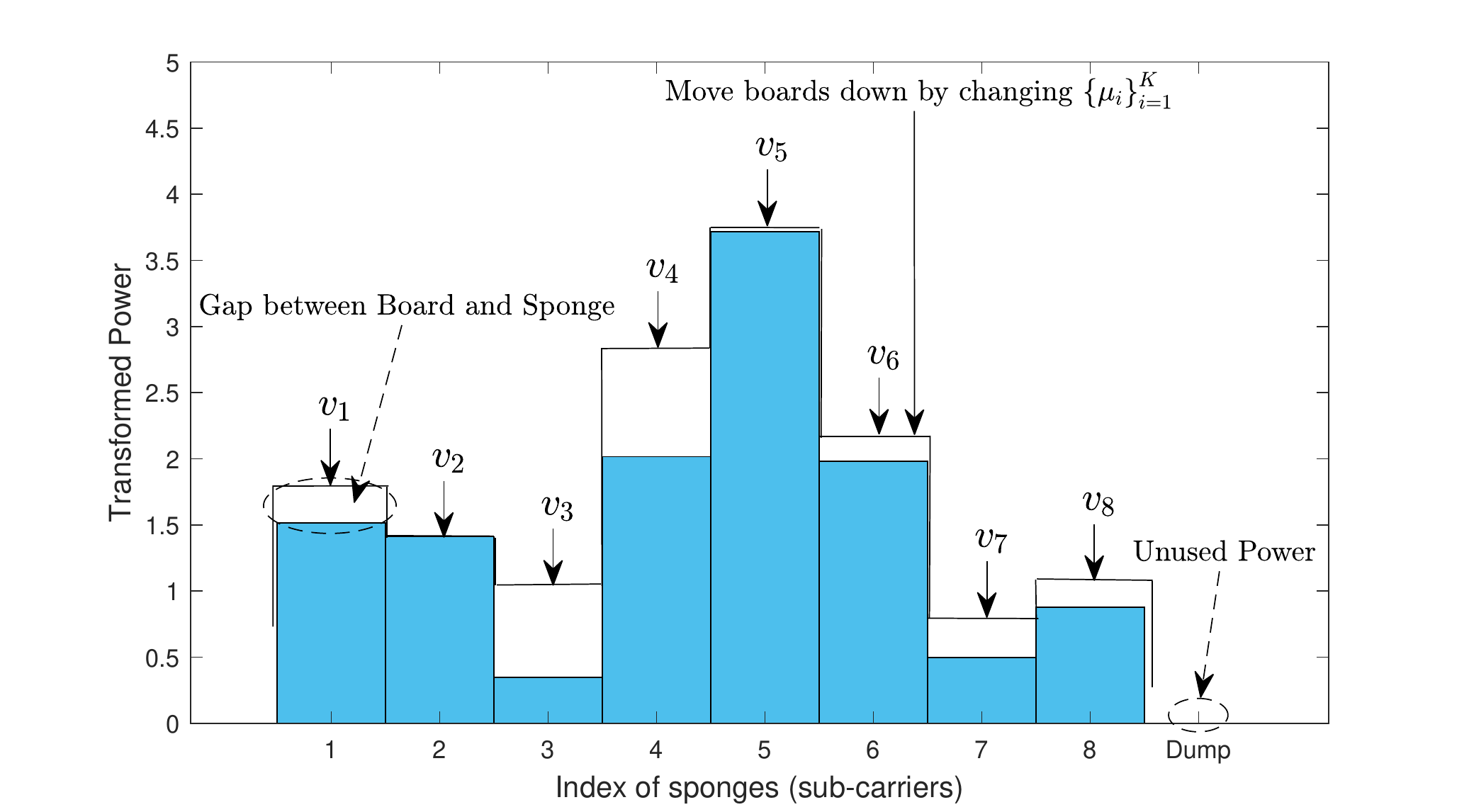}
		\label{fig:ss2}
	}
	\hfil
	\subfloat[Termination phase]{
		\hspace*{-0.6cm}\includegraphics[scale=0.6]{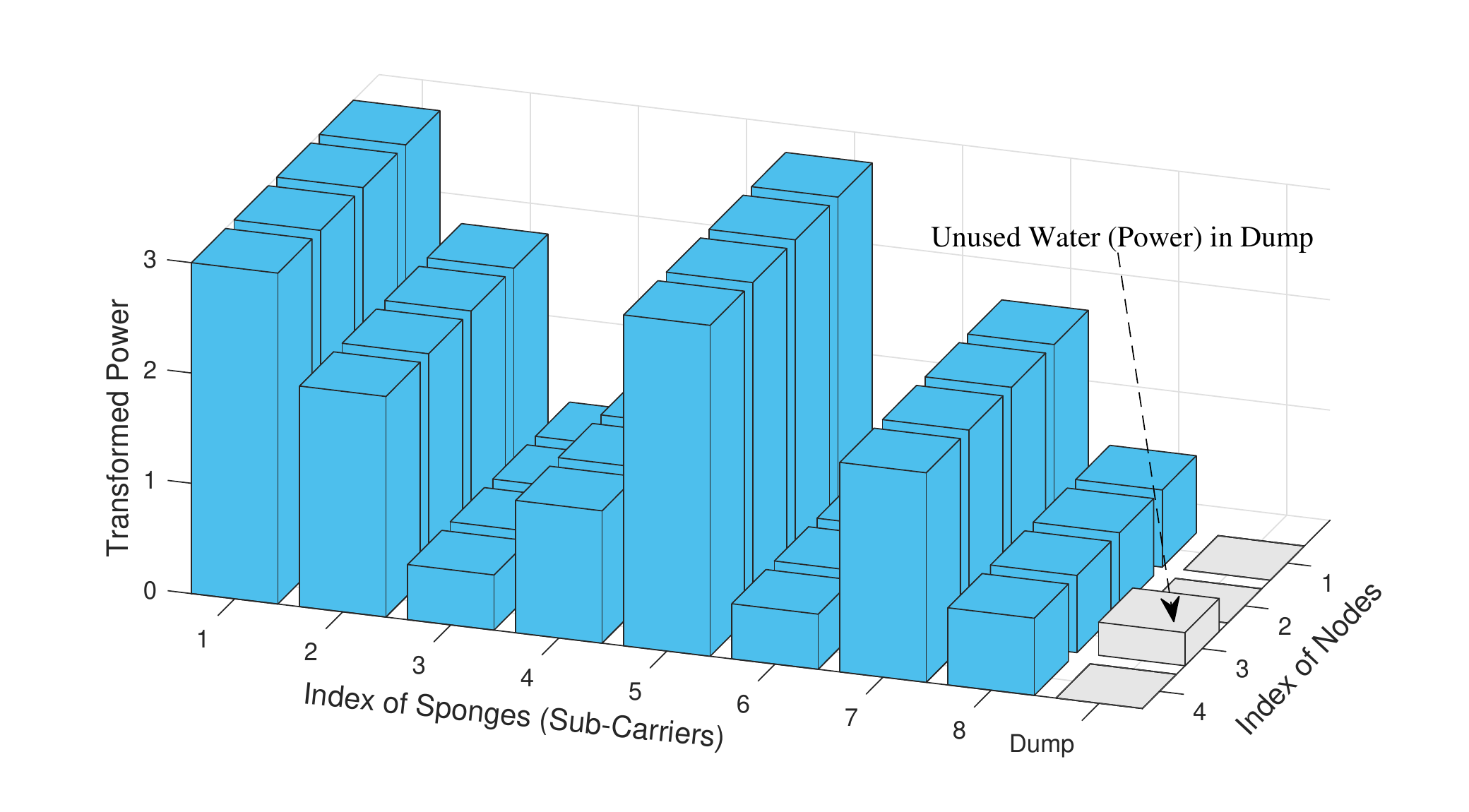}
		\label{fig:ss3}
	}
	\caption{Schematic diagram for sponge-squeezing}
\end{figure}
\subsubsection{Sponge-Squeezing}
According to the optimal solution proposed in Eqs.~\eqref{OPeta}, \eqref{Vg} and \eqref{OPpower}, the solution is quite different from the solution of the conventional power allocation with OFDM, since the solution to our problem is determined by the dot product of the Lagrange multiplier vector and the channel gain vector while the conventional solution is determined by the single Lagrange multiplier or the sum of the Lagrange multiplier vector. It means that the classical water-filling algorithm cannot be used to solve this problem. Furthermore, since the above optimization problem is a convex optimization problem, the optimal solution can be calculated by some math-tools such as CVX and MATLAB through gradient descent or interior point method. However, the performance of this method decreases rapidly as the number of sub-carriers $ N $ and the number of nodes $ K $ increase. Thus, we propose an algorithm called sponge-squeezing based on the KKT optimality conditions. It can achieve the same accuracy as CVX and improves the performance.

The principle of the algorithm, sponge-squeezing, can be summarized as follows. For any node $ i $, it owns $ N $ sponges which stand for $ N $ sub-carriers. And the volume of water that a node owns is $ P $, which stands for the power of the node $ i $. Each sponge can swell and become higher after absorbing water. Moreover, different sponge has different swell factor depending on $ \frac{1}{G_{i,g}[m]\omega_{i,g}[m]} $ (See Eq.~\eqref{OPpower}) for the node $ i $ in the $ g $-th sponge. Then, put these sponges into the container $ i $.

\emph{Initialization phase}. Node $ i $ pours water into the container, and each sponge can evenly absorb water of which the volume is $ \frac{P}{N} $. Because of the different swell factors, each sponge reach different height after absorbing water. we put $ N $ thin boards above these containers and the height of the $ g $-th thin board depends on $ v_g$ (See Eq.~\eqref{Vg}). As shown in Fig.~\ref{fig:ss1}, a node has $ 8 $ sponges, different sponge reaches different height after absorbing the same volume of water. Meanwhile, there are $ 8 $ thin boards over these sponges with different heights.

\emph{Squeezing phase}. We change the value of these Lagrange
multipliers $ \left\lbrace \mu_i \right\rbrace_{i=1}^K$, and all these thin boards start to fall and reach new heights. Note that some boards may fall fast and others fall slow due to $ v_g $. When a thin board presses on the sponge, some water in this sponge is squeezed out and is absorbed by other sponges in this container. And the lower the thin board is, the more water is squeezed out. If all these sponges are pressed under these boards, the unused water is put into the dump.

\emph{Termination phase}. We move these boards down until there is no gap between these boards and every sponge for every node while at least one dump is empty to satisfy Eq.~\eqref{Constrain}. The result with 4 nodes is shown in Fig.~\ref{fig:ss3}.

\section{Simulation Results and Discussions}
\label{Numerical Results}
In this section, we provide some simulation results of the computation rate under sub-function allocation based on CoMAC-OFDM and compare it with the conventional CoMAC schemes. In our simulation, the average signal-to-noise ratio (SNR) is the same as $ P $ because the variance of noise is set as one. For easy presentation, the abbreviation for sub-function allocation and optimal power allocation is SFA and OPA respectively.

\begin{figure}
	\centering
	\includegraphics[scale=0.7]{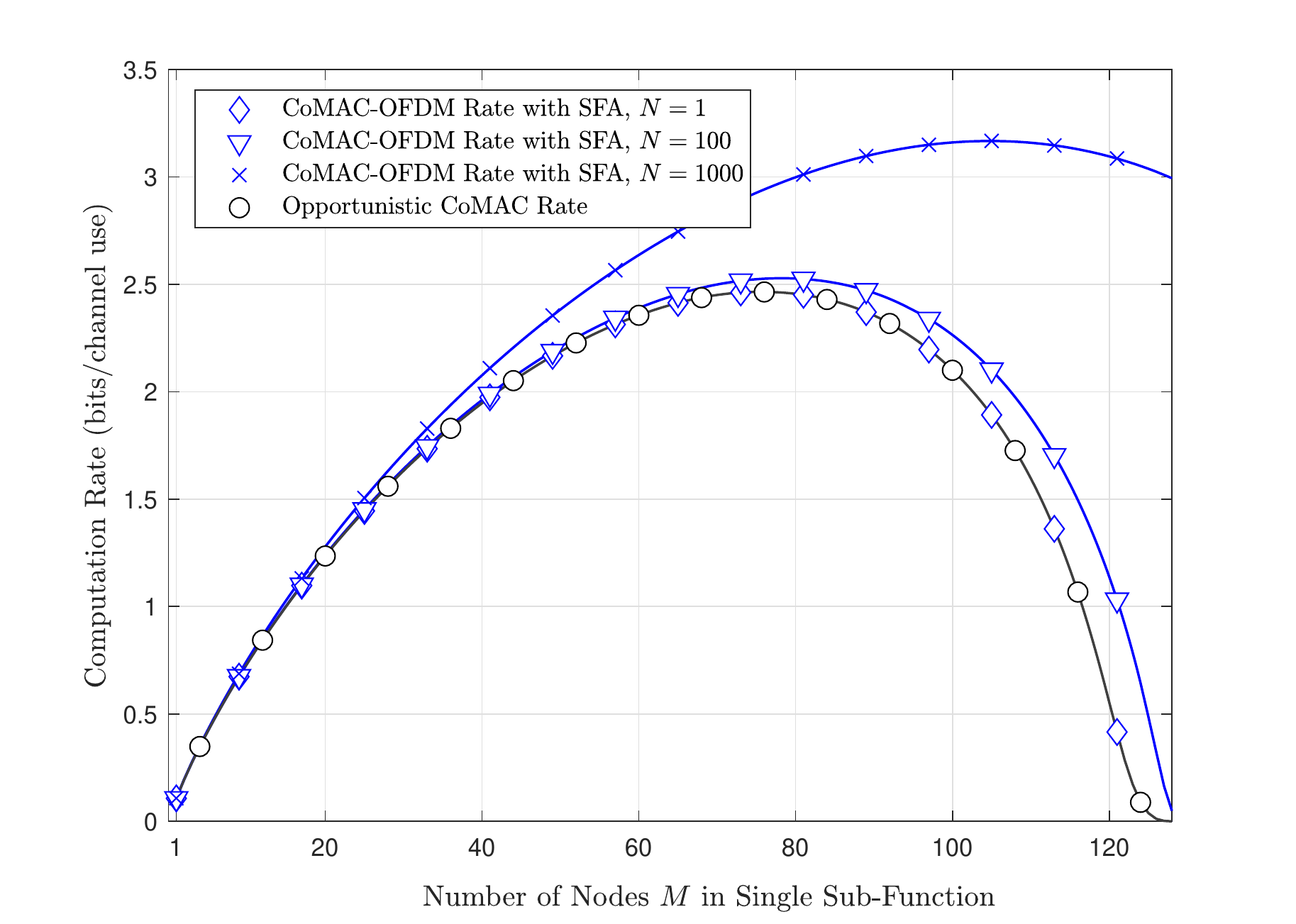}
	\caption{Computation rates with respect to $ M $ and $ N $ when $ K=128, P=10\ \rm{dB} $.}
	\label{fig:cs2}
\end{figure}

With average power constraint, CoMAC-OFDM rate with SFA in Corollary \ref{ESubPowerINC} with respect to $ M $ and $ N $ is shown in Fig.~\ref{fig:cs2}. Under sub-function allocation, $ M $ nodes participate in the computation of a sub-function. As the number of sub-carriers $ N $ increases, the computation rate increases. The opportunistic CoMAC rate in Theorem \ref{Older} is a special case when $ N=1 $ in CoMAC-OFDM rate with SFA. Furthermore, it shows that there is an optimal $ M $ to achieve the largest computation rate with a fixed $ N $, which means the computation rate is also determined by the number of sub-functions $ B $ because a desired function is divided into $ B $ sub-functions which are computed by $ M $ nodes.

\begin{figure}
	\centering
	\includegraphics[scale=0.7]{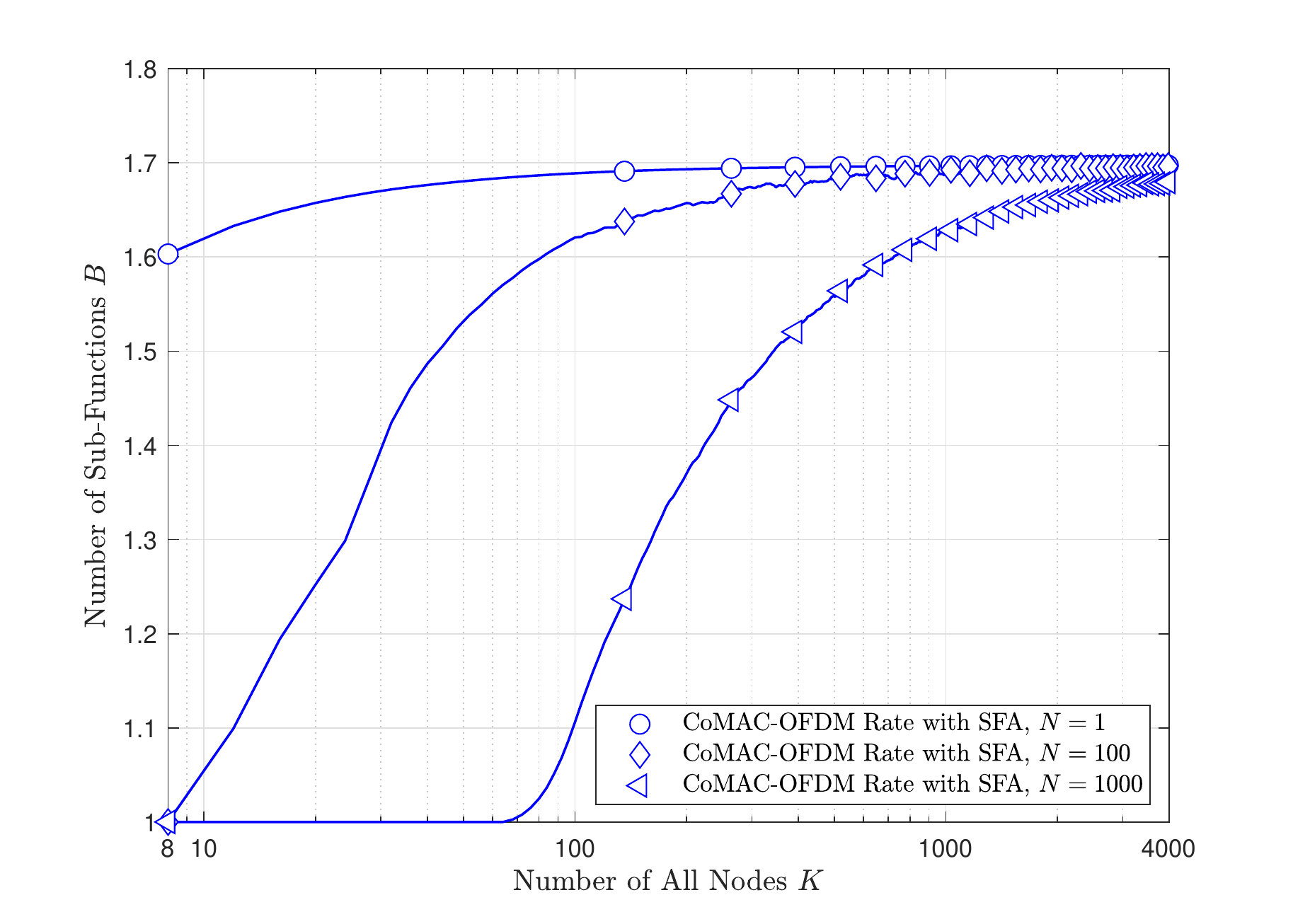}
	\caption{Optimal number of sub-functions with respect to $ K $ and $ N $ when $ P=10\ \rm{dB} $.}
	\label{fig:cs5}
\end{figure}

Since the number of sub-functions $ B $ determines the computation rate, it is necessary to find the optimal $ B $. In Fig.~\ref{fig:cs5}, the relationship between the number of nodes $ K $ and the number of sub-functions $ B $ is given. When $ K $ is small, we find that there is no need to divide the desired function into several sub-functions and allocate the desired function into each sub-carrier directly. However, for a large $ K $ (i.e, 4000), the number of sub-functions $ B $ increases and converges to $ 2 $ because $ B\in\mathbb{N} $. The result demonstrates that sub-function allocation is necessary to approach the largest computation rate.

\begin{figure}
	\centering
	\includegraphics[scale=0.7]{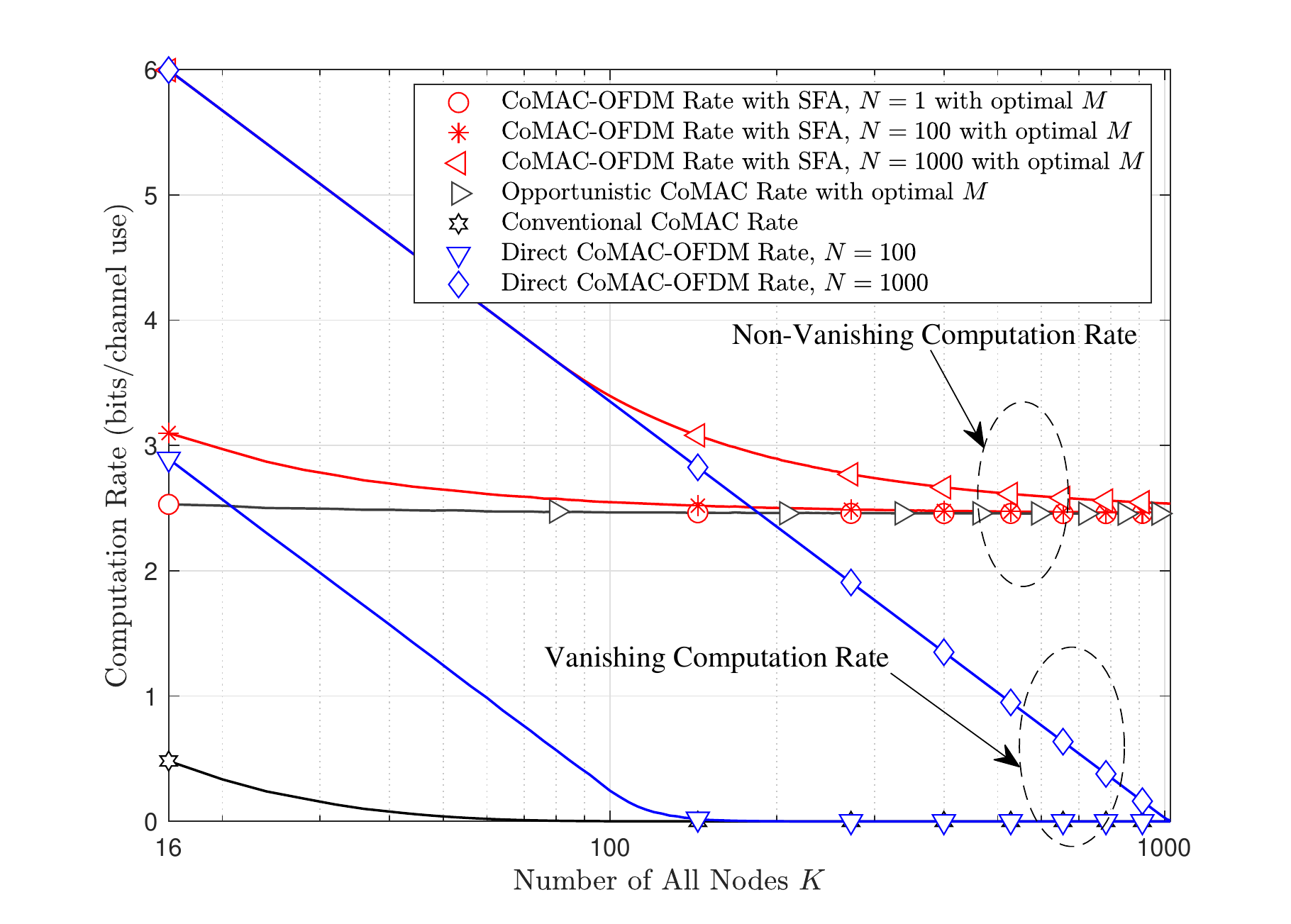}
	\caption{Comparison between CoMAC-OFDM with SFA and some conventional schemes when $ P=10\ \rm{dB} $.}
	\label{fig:cs3}
\end{figure}

In addition, we compare the direct CoMAC-OFDM rate in Corollary \ref{ComOFDMINC} and  the CoMAC-OFDM rate with SFA in Corollary \ref{ESubPowerINC} with the conventional schemes, i.e., conventional CoMAC rate and opportunistic CoMAC rate in Theorems \ref{Old} and \ref{Older} under average power constraint. As shown in Fig.~\ref{fig:cs3}, conventional CoMAC rate is the worst one and converges to 0 rapidly as $ K $ increases. Direct CoMAC-OFDM rate is improved as $ N $ increases compared with conventional CoMAC rate. However, it still cannot avoid a vanishing computation rate as $ K $ increases. Moreover, the result shows that both CoMAC-OFDM rate with SFA and opportunistic CoMAC rate can provide a non-vanishing computation rate as $ K $ increases, but CoMAC-OFDM rate with SFA can achieve a higher computation rate as $ N $ increase. It means that CoMAC-OFDM with SFA not only improves the computation rate but also provides a non-vanishing computation rate which is higher than opportunistic CoMAC rate. Furthermore, opportunistic CoMAC rate is a special case by setting $ N=1 $ in CoMAC-OFDM rate with SFA.

\begin{figure}
	\centering
	\includegraphics[scale=0.7]{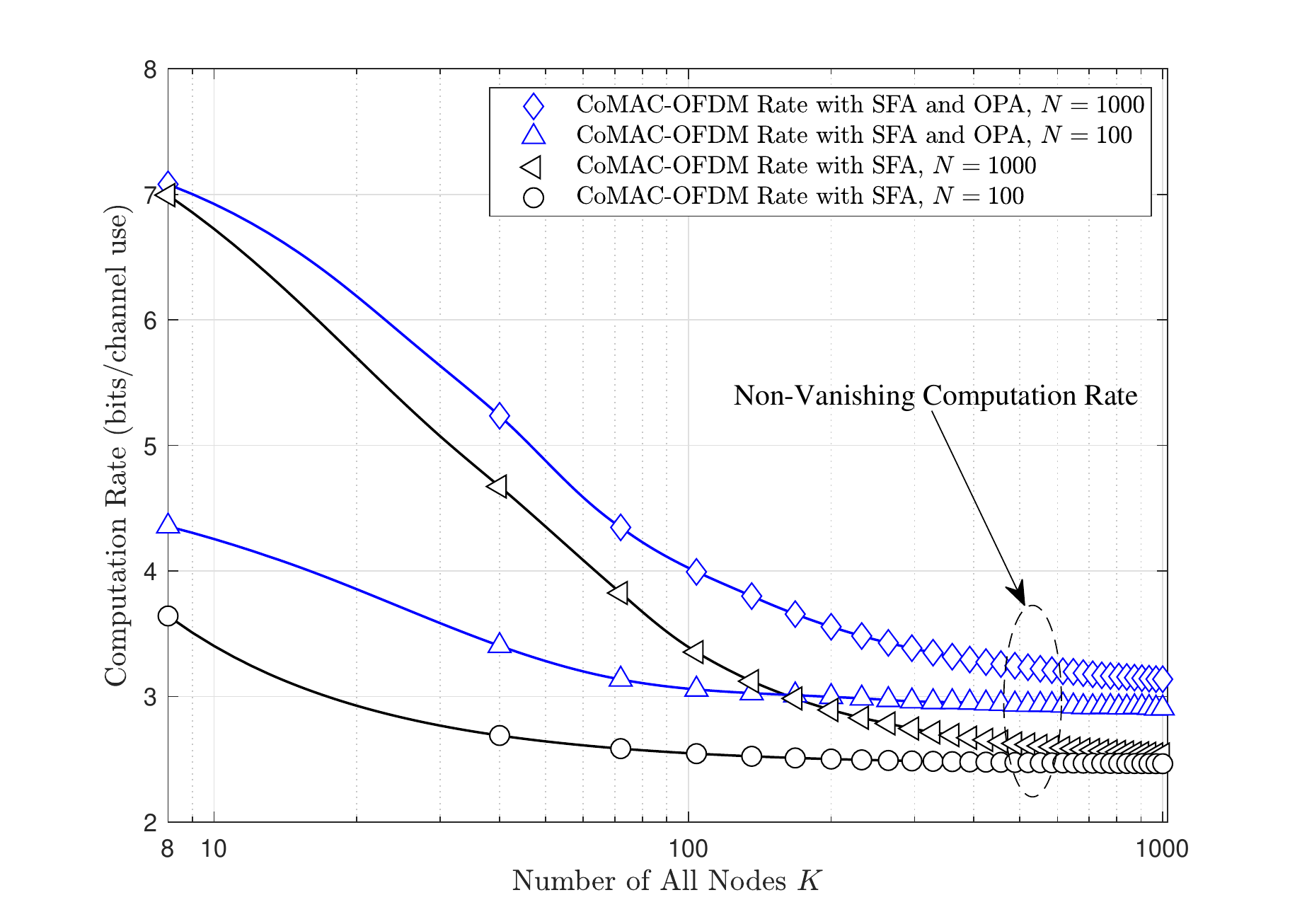}
	\caption{Comparison between optimal power allocation and average power allocation in CoMAC-OFDM with sub-function allocation when $ P=10\ \rm{dB} $.}
	\label{fig:cs4}
\end{figure}

Furthermore, the computation rate can be improved by optimal power allocation instead of average power allocation for CoMAC-OFDM. In Fig.~\ref{fig:cs4}, CoMAC-OFDM rate with SFA and OPA provide a higher rate than CoMAC-OFDM rate with SFA. It shows that the computation rate can be further improved by optimal power allocation and also provide a higher non-vanishing computation rate as $ K $ increases.

\section{Conclusion}
\label{Conclusion}
In this paper, we have studied CoMAC for wide-band transmission, and focused on designing a framework of CoMAC that can combat frequency selective fading and vanishing computation rate. To tackle these problems, we have utilized OFDM in CoMAC. CoMAC-OFDM cannot execute conventional bit allocation in different sub-carriers, because CoMAC-OFDM aims at transmitting a desired function over the air instead of bit sequences. Hence, a novel sub-function allocation has been proposed to handle this issue through the division, allocation and reconstruction of function. The theoretical expression of achievable computation rate has been derived based on the classical results of nested lattice coding, which provides a non-vanishing computation rate as the number of nodes increases. Furthermore, we have discussed an optimization problem, and a sponge-squeezing algorithm extended from the classical water-filling algorithm is proposed to carry out the optimal power allocation method. The results have showed that the computation rate can be improved by optimal power allocation and also provide a higher non-vanishing computation rate as $ K $ increases.

%
%
%
%
%
%

\bibliography{NOAH}
\bibliographystyle{IEEEtran}
\end{document}